\documentclass[aps,prd,onecolumn,noshowpacs,preprintnumbers,eqsecnum,nofootinbib]{revtex4} \usepackage[reqno]{amsmath}

\usepackage{epsfig}
\usepackage{array}
\usepackage{float}
\usepackage{lscape,graphicx}
\usepackage{amssymb}
\usepackage[usenames,dvipsnames]{color}
\usepackage{ulem}
\usepackage{enumerate}
\usepackage{natbib}

\newcommand{\eV}{\,{\rm eV}}

\newcommand{\GeV}{\,{\rm GeV}}
\newcommand{\TeV}{\,{\rm TeV}}

\newcommand{\etc}{\,{\rm ...} }

\newcommand{\ie}{{\it i.e.}}

\newcommand{\beq}{\begin{equation}}
\newcommand{\eeq}{\end{equation}}
\newcommand{\bea}{\begin{eqnarray}}
\newcommand{\eea}{\end{eqnarray}}
\newcommand{\ba}{\begin{array}}
\newcommand{\ea}{\end{array}}

\newcommand{\ovl}{\overline}
\newcommand{\eps}{\epsilon}
\newcommand{\al}{\alpha}
\newcommand{\be}{\beta}
\newcommand{\lam}{\lambda}
\newcommand{\D}{\Delta}

\newcommand{\Fs}{\mathcal{F}}
\newcommand{\mup}{\mu^{\prime}}
\newcommand{\mupp}{\mu^{\prime\prime}}
\newcommand{\K}{\kappa}
\newcommand{\ii}{ {\rm i} }
\newcommand{\dmsol}{\mbox{$\Delta m^2_{\odot}$}}
\newcommand{\dma}{\mbox{$\Delta m^2_{\rm A}$}}

\newcommand{\Yn}{y_{N_3} }
\newcommand{\YDn}{ y_{\Delta N_D}}
\newcommand{\YDl}{ y_{\Delta \ell }}
\newcommand{\YDs}{ y_{\Delta S}}
\newcommand{\YDhc}{y_{\Delta H_3 }}
\newcommand{\YDha}{y_{\Delta H_1 }}
\newcommand{\YDhb}{ y_{\Delta H_2 }}


\begin{document}


\preprint{{\it CFTP/11-015}}

\title{A Common Framework  for Dark Matter,  Leptogenesis  and Neutrino Masses}

\author{Fran\c cois-Xavier~Josse-Michaux} \email{fxjossemichaux@gmail.com}
\author{Emiliano Molinaro} \email{emiliano.molinaro@ist.utl.pt}

\affiliation{Centro de F\'isica Te\'orica de Part\'iculas (CFTP), Instituto Superior T\'ecnico,\\
Technical University of Lisbon, 1049-001 Lisboa, Portugal}

\begin{abstract}
We study  a seesaw-type extension of the Standard Model
in which the symmetry group is enlarged by a global $U(1)$.  
We introduce  adequate scalar and fermion representations which naturally explain the smallness of neutrino masses. 
With the addition of a viable scalar  Dark Matter candidate, an original scenario of leptogenesis emerges.
We solve the relevant set of Boltzmann equations and show how leptogenesis can be
successfully implemented at the TeV scale.
The constraints on the scalar mass spectrum are derived and the Dark Matter phenomenology is discussed.
\end{abstract}

\maketitle
\tableofcontents

\section{Introduction}

Now that we entered in the LHC era, 
the Standard Model (SM) of elementary particles can be definitively tested. Until now, the SM has been extremely successful, as no strong signals of new physics have been observed so far at particle accelerators.
However other experiments have long-time evidences for the need of extensions of the SM particle content. Neutrino oscillations are the prime among them on the particle side, but the compelling gravitational evidences for the existence of Dark Matter (DM), as well as the observation of a matter-antimatter asymmetry in the Universe all call for new physics.

From neutrino oscillation experiments 
we know that at least two neutrinos should be massive with an
overall mass scale $m_{\nu}$ constrained by different observations: $m_{\nu}\lesssim 1$ eV. 
More precisely, experiments with solar, atmospheric, reactor and accelerator neutrinos \cite{neutrino1}-\cite{neutrino10}
set two  mass scales in the theory, $\dmsol$ and $\dma$, which drive the solar and atmospheric neutrino
oscillations, respectively \cite{PDG10}:
\bea
		\dmsol\;=\left(7.59^{+0.20}_{-0.21}\right)\times 10^{-5}\,\eV^{2}\;\,,\quad\quad\dma\;=\;\left(2.43\pm 0.13\right)\times 10^{-3}\,\eV^{2}\,.\label{numassexp}
\eea
Moreover, these experiments show that flavor neutrino mixing, described in terms of the PMNS  \cite{BPont57,MNS62,BPont67} matrix,
is characterized by two large mixing angles, $\theta_{12}$ and $\theta_{23}$, and a small one, $\theta_{13}$~\cite{neutrino11}.

On the cosmological side,
the matter content of the Universe has been measured with precision by WMAP \cite{WMAP}.
The resulting Dark Matter and baryon number densities, 
$\Omega_{\rm DM}$ and $\Omega_{\rm B}$, are
\bea \label{Omegas}
\Omega_{\rm DM}= 0.229 \pm 0.015 \,,\quad \Omega_{\rm B}= 0.0458 \pm 0.0016\,.
\eea
Several gravitational observations confirm the existence of non-baryonic matter \cite{DMevidences}, which is not accounted for in the SM. New  physics extensions are then necessary and various viable DM candidates exist~\cite{DMevidences}. However, the real nature of DM is still elusive, as no direct proof has been observed - or firmly confirmed - so far \cite{DMdd1}-\cite{XENON100}. 
The measurement by WMAP of the baryonic matter content of the Universe is in agreement with the value predicted by Big-Bang Nucleosynthesis from the observations of the primordial abundances \cite{BBN}. However, an excess of baryons over antibaryons is observed, and the standard cosmological scenario fails to  explain this Baryon Asymmetry of the Universe (BAU). Particle physics extensions of the SM are advocated to justify this: in relation with neutrino masses, the leptogenesis scenario \cite{lepto,luty} constitutes one of the most elegant solutions. 

In this  paper we study a minimal extension of the SM
in which it is possible to address, in a consistent way, the three puzzles
listed above. The model is based on a global $U(1)_{B-\tilde{L}}$ symmetry, 
which is spontaneously broken below the electroweak symmetry
breaking (EWSB) scale. The $\tilde{L}$ charge is a generalization of the usual lepton number $L$, as $\tilde{L}=L$ for the SM particles.
The light neutrino masses are explained within a seesaw framework~\cite{SeesawI}, through the introduction of a SM singlet Dirac fermion $N_D$, together with three Brout-Englert-Higgs scalar particles: two $SU(2)_{W}$ doublets $H_{1,2}$ and a SM singlet $H_3$, which drive the EWSB by acquiring non-zero vacuum expectation values (vevs). All these extra degrees of freedom are charged under the global $U(1)_{B-\tilde{L}}$ symmetry. 
In \textit{e.g.}~\cite{MNUothers}, neutrino masses were generated in models with similar scalar spectrum and/or based on a (spontaneously broken) global symmetry, although in different physical frameworks.
In our scenario, when the seesaw scale is set in the TeV-range, such a particle content provides a UV-completion of the inverse-seesaw mechanism of neutrino mass generation~\cite{Mohapatra:1986bd}.

Nevertheless, with just this particle content, neither the observed amount of baryon asymmetry nor the Dark Matter abundance, eq. (\ref{Omegas}), can be accounted for.

In order to solve also these two important issues, we complete the model
by introducing a Majorana neutrino $N_{3}$ and a complex scalar $S$.
Both particles are SM singlets, although $S$ is charged under the global $U(1)_{B-\tilde{L}}$. 
The particle content of the model is summarized in Tab.~\ref{FieldAssign}, together with the $U(1)_{B-\tilde{L}}$ quantum numbers of the fields.
The new scalar $S$ provides, after the breaking of  $U(1)_{B-\tilde{L}}$, a natural Dark Matter candidate, whose stability is guaranteed by a remnant $\mathcal{Z}_{2}$ symmetry.

It is remarkable that the introduction of $S$ allows a TeV scale scenario of leptogenesis.
Indeed, as  the Majorana field $N_{3}$ couples to $N_{D}$ and $S$, the out-of-equilibrium $CP$-violating decays of $N_{3}$ can generate a number density asymmetry in $N_{D}$ and $S$, resembling the standard thermal leptogenesis mechanism in the type I seesaw extension of the SM.
However, in the present case leptogenesis is implemented in two steps: first an asymmetry in $N_{D}$ and $S$ is generated by the decays of $N_{3}$; 
in a second phase, the Dirac neutrino asymmetry is transferred to SM leptons by sufficiently fast neutrino Yukawa interactions. The latter set a link between successful leptogenesis 
and viable neutrino mass generation via the seesaw mechanism.
Finally, as in standard leptogenesis,  non-perturbative sphaleron effects partly convert this lepton asymmetry into a net baryon number \cite{KRS}.

\begin{table}[t]
\begin{center}
\begin{tabular}{|c||c|c|c|c||c|c|c||c|}
\hline \hline
\rule[0.15in]{0cm}{0cm}{\tt Field}& $\ell_{\alpha}$ & $e_{R\alpha}$ & $N_{D}$ & $N_{3}$ & $H_{1}$ & $H_{2}$ & 
$H_{3}$ & $S$  \\
\hline
\rule[0.25in]{0cm}{0cm}$B-\tilde{L}$ & -$1$ & -$1$ & -$1$ & $0$ & $0$ & $2$ & 
-$2$ & -$1$
\\\hline\hline
\end{tabular}
\caption{Charge assignment of the fields.}\label{FieldAssign}
\end{center}
\end{table}%

In Section~\ref{neutrsec} we discuss  neutrino mass generation through the (inverse) seesaw 
mechanism. In the subsequent Section~\ref{2stepslep} we tackle the problem of the BAU and 
study the constraints on the parameter-space of the model imposed
by  successful leptogenesis. The computation of the $CP$ asymmetry and the set of coupled Boltzmann equations governing the number density evolutions are reported in the final appendices.
In Section~\ref{ScalarSect} we discuss the scalar sector of the theory,
deriving the  mass spectrum and  corresponding constraints.
In Section~\ref{DMSec} we  study  the possibility of having a viable 
Dark Matter in the model and comment  on the possible observation of DM in direct detection experiments. 
Finally, in the last section we summarize the main results of the paper.

\mathversion{bold}
\section{Neutrino Masses with a Global $U(1)_{B-\tilde{L}}$}\label{neutrsec}
\mathversion{normal}

An effective Majorana neutrino mass term is generated below the EWSB scale from the following part of the interaction Lagrangian:
\begin{eqnarray}
	-\mathcal{L}_{\rm int} \supset\,M\,\overline{N}_{D}N_{D}\,+\left( y_{1}^{i}\,\overline{N}_{D}\,\widetilde{H}_{1}^{\dagger}\,\ell_{i}\,
	+\,y_{2}^{j}\,\overline{N}_{D}^{c}\,\widetilde{H}_{2}^{\dagger}\,\ell_{j}
	\,+\,\frac{\alpha}{\sqrt{2}}\,H_{3}\,\overline{N}_{D}\,N_{D}^{c}\,+\,{\rm h.c.}\right)\,,		
\label{LYuk}
\end{eqnarray}
where $\ell_{i}=(\nu_{i\,L},e_{i L})^{T}$ ($i=e,\mu,\tau$), $N_{D}^{c}\equiv C\overline{N}_{D}^{T}$ 
and $\widetilde{H}_{k}\equiv -i\sigma_{2}H_{k}^{*}$ ($k=1,2$).~\footnote{
$C$ is the usual charge conjugation matrix of Dirac spinors.}~The coupling constant
$\alpha$ and the neutrino Yukawa couplings $y_{1,2}^{i}$ are complex parameters.
As we will see in the following, the phase of $\alpha$ plays a crucial role in the
generation of the $CP$ asymmetry necessary for the production of the observed amount of BAU.

The terms reported in the Lagrangian  (\ref{LYuk}) provide a dynamical realization of
the inverse seesaw mechanism \cite{Mohapatra:1986bd} for the generation of neutrino masses 
in the case the mass of the Dirac field $N_{D}$ is taken in the TeV-range.
More specifically, in our scenario the standard lepton charge $L$ is explicitly violated by the interactions 
involving the couplings $y_{2}^{i}$ and $\alpha$. 
Consequently, we expect that the active neutrino masses, generated through the (inverse) seesaw mechanism, 
do directly depend on these parameters. 
The model, in this minimal form, predicts two massive and one massless active neutrinos.

The seesaw mass scale $M$ is  a free parameter of the theory
and can assume arbitrarily large values above the EWSB scale. However,
in the following we will be mostly interested in the case where $M$ is taken at the TeV scale.
At energies much smaller than $M$, $N_{D}$ is integrated out and 
we get at second order in $1/M$ the $(B-\tilde{L})$-conserving effective 
Lagrangian:~\footnote{We do not include flavor kinetic mixing terms in the Lagrangian (\ref{Lageff}), which arise by 
dimension 6 effective fermion operators.}
\begin{eqnarray}
	-\mathcal{L}_{\rm eff}&\supset&-\;\frac{y_{1}^{i}y_{2}^{j}+y_{1}^{j}y_{2}^{i}}
	{2M}\left( \overline{\ell}^{c}_{j}\,\widetilde{H}_{2}^{*} \right)
	\left(\widetilde{H}_{1}^{\dagger}\,\ell_{i}\right)\;+\;
	\frac{y_{1}^{i}y_{1}^{j}\alpha^{*}}{\sqrt{2}M^{2}}\left( \overline{\ell}^{c}_{j}\,\widetilde{H}_{1}^{*} \right)
	\left(\widetilde{H}_{1}^{\dagger}\,\ell_{i}\right)\,H_{3}^{*}\nonumber\\\label{Lageff}\\
	&& +\;
	\frac{y_{2}^{i}y_{2}^{j}\alpha}{\sqrt{2}M^{2}}\left( \overline{\ell}^{c}_{j}\,\widetilde{H}_{2}^{*} \right)
	\left(\widetilde{H}_{2}^{\dagger}\,\ell_{i}\right)\,H_{3}\;+\;{\rm h.c.}\,,	\nonumber
\end{eqnarray}
where the sum over the flavor indices $i$ and $j$ is understood. 
When the neutral components of the scalar fields  $H_{k}$ ($k=1,2,3$)
take a non-zero vev,
the operators in (\ref{Lageff}) generate a Majorana mass term
for the flavor neutrino fields $\nu_{i\, L}$. Indeed,   
taking $\langle H_{i} \rangle=\left(0,\,v_{i}/\sqrt{2}\right)^{T}$ ($i=1,2$) and $\langle H_{3}\rangle=v_{3}/\sqrt{2}$ in (\ref{Lageff}),
we obtain the neutrino mass Lagrangian
\begin{equation}
	\mathcal{L}_{m_{\nu}}\;=\; -\frac{1}{2}\overline{{\bf \nu^{c}_{R}}}\,m_{\nu}\,{\bf\nu_{L}}+{\rm h.c.}\,,
\end{equation} 
where ${\bf\nu_{L}}\equiv(\nu_{eL},\nu_{\mu L},\nu_{\tau L})$, ${\bf\nu^{c}_{R}}\equiv C{\bf \overline{\nu}_{L}}^{T}$ and
\begin{eqnarray}
	\left(m_{\nu}\right)_{ij} &=& -\left(y_{1}^{i}\,y_{2}^{j}+y_{2}^{i}\,y_{1}^{j}
	-y_{1}^{i}\, y_{1}^{j}\,\alpha^{*}\,\frac{v_{1}\,v_{3}}{v_{2}\,M}-
	y_{2}^{i}\, y_{2}^{j}\,\alpha\,\frac{v_{2}\,v_{3}}{v_{1}\,M}
	\right)\frac{v_{1}\, v_{2}}{2\,M}\,.
\end{eqnarray}
The masses of the two active neutrinos are given by
\begin{eqnarray}\label{nueig}
	m_{\pm} &\simeq& \frac{1}{4}\left|v_{3}\frac{v_{2}^{2}}{M^{2}}y_{2}^{2}\alpha+v_{3} \frac{v_{1}^{2}}{M^{2}}y^{2}_{1}\alpha^{*}
	-2v_{1}\frac{v_{2}}{M}y_{12}\pm\sqrt{\left(v_{3}\frac{v_{2}^{2}}{M^{2}}y_{2}^{2}\alpha+v_{3} \frac{v_{1}^{2}}{M^{2}}y_{1}^{2}\alpha^{*}
	-2v_{1}\frac{v_{2}}{M}\,y_{12}\right)^{2}+4\,v_{1}^{2}\frac{v_{2}^{2}}{M^{2}}\eta_{12}^{2} } \right|,
\end{eqnarray}
where we define $y_{12}=y_{1}^{e}y_{2}^{e}+y_{1}^{\mu}y_{2}^{\mu}+y_{1}^{\tau}y_{2}^{\tau}$,
$y_{k}=\sqrt{(y_{k}^{e})^{2}+(y_{k}^{\mu})^{2}+(y_{k}^{\tau})^{2}}$ ($k=1,2$) and $\eta_{12}=\sqrt{(y_1^{e}y_{2}^{\mu}-y_{2}^{e}y_{1}^{\mu})^{2}+
(y_1^{e}y_{2}^{\tau}-y_{2}^{e}y_{1}^{\tau})^{2}+(y_1^{\mu}y_{2}^{\tau}-y_{2}^{\mu}y_{1}^{\tau})^{2}}$.
As usual in two-Higgs doublet models, the vevs of the two scalar doublets, $v_{1}$ and $v_{2}$, 
are related to the EWSB scale: $\sqrt{v_{1}^{2}+v_{2}^{2}}\equiv v\simeq 246$ GeV.
As explained in Section \ref{ScalarSect}, the hierarchy among the Higgs vevs is tightly constrained in our model, in particular from the presence of a massless Goldstone boson associated with the spontaneous breaking of the global $U(1)_{B-\tilde{L}}$: phenomenological constraints enforce $v_{2}\ll v_{1,3}$, and by convention we impose $v_{3}\leq v$. As we will see in Section \ref{ScalarSect}, this hierarchical pattern is easily realized in the model. Typically, for $|\alpha|\approx 0.01$,  $M\approx 1$ TeV
and a scalar spectrum with $v_{2}\approx 10$ MeV, $v_{3}\approx 100$ GeV, the neutrino Yukawa couplings are $|y_{1,2}|\approx 10^{-4}$.
 
 The Yukawa interaction $\al \,H_{3}\,\overline{N}_{D}\,N_{D}^{c}$ generates after EWSB a small Majorana mass term
 for the two chiral components of the  Dirac field $N_D$, 
 which is then split into two quasi-degenerate Majorana fermions:
they behave as a pseudo-Dirac pair \cite{Wolfenstein}-\cite{Branco}, with a mass difference of the order $2\,v_{3}\,|\alpha|$.
Such scenarios have been studied in detail in \cite{delAguila}, where it was shown that a high-level of degeneracy prevents the Majorana nature of these states to be observed at colliders,
LHC included. Indirect signals of TeV scale pseudo-Dirac neutrinos coupled to charged leptons can in principle be
observed both in lepton flavor violating processes, $e.g.$ charged lepton radiative decays $\ell_{i}\to \ell_{j}\,\gamma$
and  $\mu-e$ conversion in nuclei, and in experiments searching for lepton number violation, such as neutrinoless double beta decay processes. 
For these processes, the contribution of the heavy neutrinos to the decay rate may be relevant/dominant  in the case
of $M\approx (100-1000)$ GeV, $|\alpha| v_{3}/M\approx 10^{-3}-10^{-2}$ and for sizable neutrino Yukawa couplings, $|y_{1,2}|\approx 10^{-2}$~\cite{Molinaro}.

Finally, we remark that the coupling $\alpha$ is not strictly required in order to obtain two massive neutrinos, 
whereas the introduction of $y_2$ is mandatory. Actually, one can show that $y_{1}$ and $y_{2}$ are also sufficient to fully reconstruct
the low-energy neutrino data, up to a normalization factor \cite{ThomasBelen}.
From eq.~(\ref{nueig}) we get the following relation:
\begin{equation}\label{eta12cond}
	\left| \eta_{12} \right|\,v_{1}\,v_{2}\,\frac{1}{M}\; =\; 2\,\left(\dmsol\,\dma\right)^{1/4}\,.
\end{equation}
This equation clearly shows that for $y_{2}=0$ or for $(y_{2}^{e},y_{2}^{\mu},y_{2}^{\tau})$ aligned with $(y_{1}^{e},y_{1}^{\mu},y_{1}^{\tau})$, $\vert \eta_{12} \vert =0$ and only one neutrino is massive, in contradiction with neutrino oscillation data. Barring accidental cancellations, eq.~(\ref{eta12cond}) implies 
\begin{equation}\label{v2low}
	\vert y_{1}\vert\,\vert y_{2}\vert\;\sim\; 2\times 10^{-8}\,\left(\frac{M}{1~\text{TeV}}\right)\,\left(\frac{10~\text{MeV}}{v_{2}}\right)\,.
\end{equation} 


\section{Two-Step Leptogenesis}\label{2stepslep}

Before discussing how the baryon asymmetry is generated in our scenario, let us briefly recall the standard picture of leptogenesis, based on the type I seesaw
extension of the Standard Model. For a detailed discussion, see \cite{Davidson:2008bu} and references therein. 
In the standard scenario, at least 2 massive right-handed (RH) neutrinos, which are $SU(2)_{W}\times U(1)_{Y}$ singlets, are introduced and
 couple to lepton doublets through Yukawa interactions.  These singlets are Majorana fermions whose mass $M_{R}$ is not
related to the electroweak scale  and can assume arbitrarily large values.
The RH neutrinos evolve together with the SM particles in a hot but expanding Universe; 
when the temperature drops down below $M_{R}$, they start to decouple and decay out-of-equilibrium in both leptons and antileptons.
If $CP$ is violated in these processes, a non-zero asymmetry is produced, which
is subsequently converted into a net baryon number by fast sphaleron interactions. 
The latter are non-perturbative effects, in thermal equilibrium above the EWSB scale up to 
temperatures  $T\lesssim 10^{12}$-$10^{13}$ GeV~\cite{SphaRelax1}. 
Several interactions should be considered for an accurate determination of the efficiency of leptogenesis in producing a baryon asymmetry.
Spectator processes play an important role in modifying the production/depletion mechanisms, most notably by spreading the lepton asymmetry into different species. 

In the present case, given the particle content and the charge assignment listed in Tab~\ref{FieldAssign}, the interaction Lagrangian receives, besides the operators of the seesaw sector in eq.~(\ref{LYuk}), contributions from the extra Majorana field $N_{3}$ and the scalar $S$:
\bea
-\mathcal{L}_{\rm int} &\supset&
		\mu_{S}^2\, S^{*} S\,+\,\frac{1}{2}M_{3}\,\overline{N}_{3}\,N_{3}^{c}\,+\left(g\,S\,\overline{N}_{D}\,N_{3}\,
		- \frac{\mupp}{\sqrt{2}} S^2 H_3^{*}\,+\,{\rm h.c.}\right) 
	\label{Lint2}\,,
\eea
where $M_{3}$, $\mupp$ and $g$ can be set real by a redefinition of the phases of $N_{3}$, $S$ and $H_{3}$. 
We impose  $N_{3}$ to be heavier than $N_{D}$ and $S$.

In this model, the generation of a baryon asymmetry proceeds in two different phases. 
In a first phase, which is similar to the standard leptogenesis scenario, an asymmetry in Dirac neutrinos $N_{D}$ and in $S$ is generated by the out-of-equilibrium decays of the Majorana field $N_{3}$. As we describe below, the $CP$ asymmetry in $N_{3}$ decays is only possible after the introduction of $S$, carrying the same $B-\tilde{L}$ quantum number as $N_{D}$.

Besides decays and inverse decays, several scatterings affect $N_{D}$ and $S$ asymmetries. All these interactions conserve the total $B-\tilde{L}$ charge.
In a second phase, owing to the neutrino Yukawa couplings, the produced $N_{D}$ and $S$ asymmetries are transferred and reprocessed into a lepton asymmetry. In this second phase, the sphaleron processes partly convert the so produced lepton asymmetry into a final baryon number, as in the standard picture.

This model can thus be viewed as the SM augmented with a second Higgs doublet, combined with a hidden sector composed of the fields $N_{3}$, $S$ and $H_{3}$. The two sectors share a conserved $B-\tilde{L}$ charge through the Dirac neutrino $N_{D}$. In that extent, the role of the neutrino Yukawa couplings  is central both in the generation of light neutrino masses and in the production of a BAU, in agreement with observations.

\begin{figure}[t!]
\begin{center}
\includegraphics[width=0.7\textwidth]{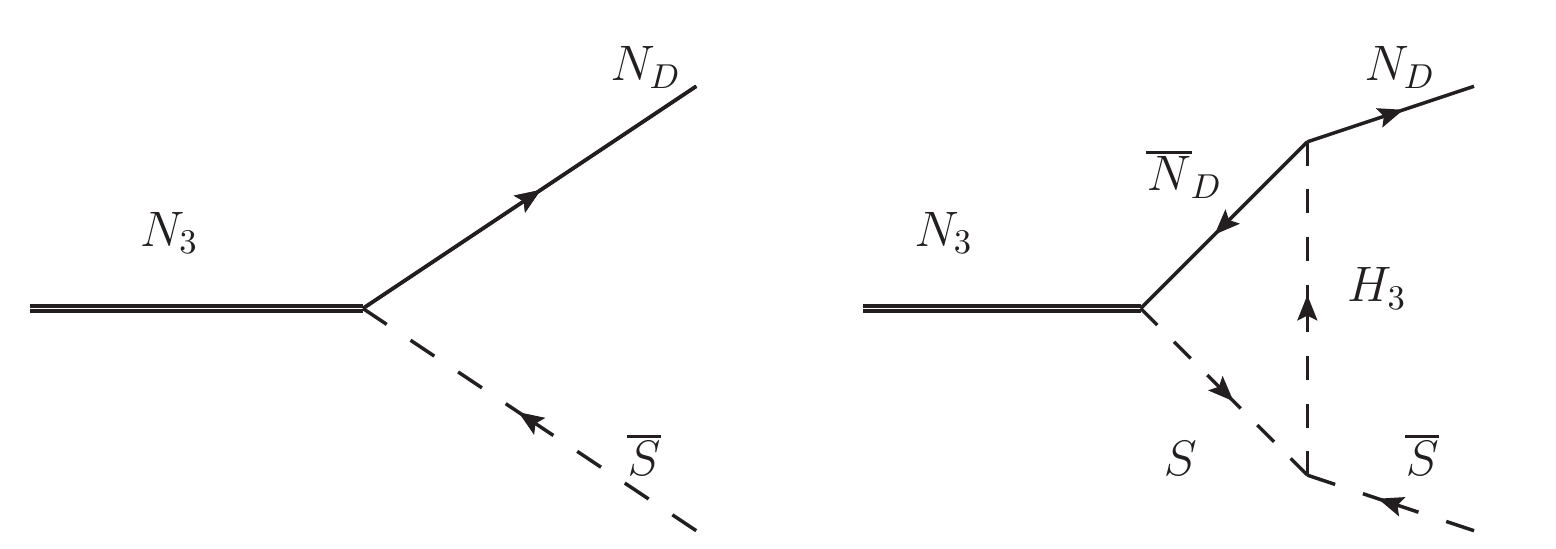}\
\caption{Diagrams contributing to the $CP$ asymmetry in the decays of $N_{3}$.}
\label{CPAsymmGraph}
\end{center}
\end{figure}

\mathversion{bold}
\subsection{The $CP$ asymmetry $\eps_{CP}$}
\mathversion{normal}
In the standard leptogenesis scenario a $CP$ asymmetry is generated by the interference between the tree-level and the one-loop corrections to the decay amplitude of the heavy Majorana neutrinos~\cite{luty,Ecp}, owing to the presence of at least two heavy states. In our case, with only one heavy neutrino $N_{D}$, no $CP$ violation is produced in $N_{D}$ decays. 
On the other hand, a non-zero $CP$ asymmetry can be generated by the addition of $N_{3}$ and $S$, from the interference between the tree-level and one-loop correction to $N_{3}$ decay amplitude, whose Feynman diagrams are depicted in Fig.~\ref{CPAsymmGraph}.

The detailed computation of the $CP$ asymmetry in $N_{3}$ decays is provided in Appendix~\ref{CompCPAsymm}.
We report below the resulting expression in the limit $M_{3}\gg M, \mu_{S}$:
\bea
\eps_{CP}&\simeq&- \frac{ {\rm Im}(\al)}{16\pi}\,\frac{\mupp}{M_3}\,.
\label{ApprECP}
\eea 
Despite of the fact that $N_{3}$ decays depend on the coupling constant $g$, the latter being a real parameter does not enter in the expression of $\eps_{CP}$, cf. eq.~(\ref{ECPexact}). The only source of $CP$ violation relevant for leptogenesis is the phase of the complex parameter $\al$  in the Lagrangian~(\ref{Lint2}).
It is remarkable that, in contrast to the standard leptogenesis scenario, 
there is no direct dependence of $\eps_{CP}$ on the neutrino Yukawa couplings $y_{1,2}$. 
Still, a connection between the leptogenesis $CP$-violating phase and the light neutrino masses exists and is actually provided by the imaginary part of $\alpha$. 
We remark that  the parameter $\mupp$ in (\ref{ApprECP}) enters in the mass splitting between the real and imaginary parts of $S$ (cf. eq.~(\ref{DMmass})) and therefore determines which is the DM candidate of the model, as shown in Section~\ref{DMSec}.
Provided $\mupp$ is not too much suppressed compared to $M_3$ and the phase of $\alpha$ is different from zero, $\eps_{CP}$ takes sizable values.
We typically have:
\bea
\eps_{CP}\simeq -2\times 10^{-6}\,\left(\frac{\mupp}{1\GeV}\right)\,\left(\frac{10\,\TeV}{M_3}\right)\,{\rm Im}(\al)\,.
\eea

\subsection{Asymmetry productions}
We discuss now the salient aspects of leptogenesis in our scenario. 
We eventually distinguish between two stages of production, but we shall emphasize that these stages are not necessarily consecutive 
and may occur in the same temperature range. 
\begin{figure}[h!]
\begin{center}
{\rm $a)$}\includegraphics[width=0.8\textwidth]{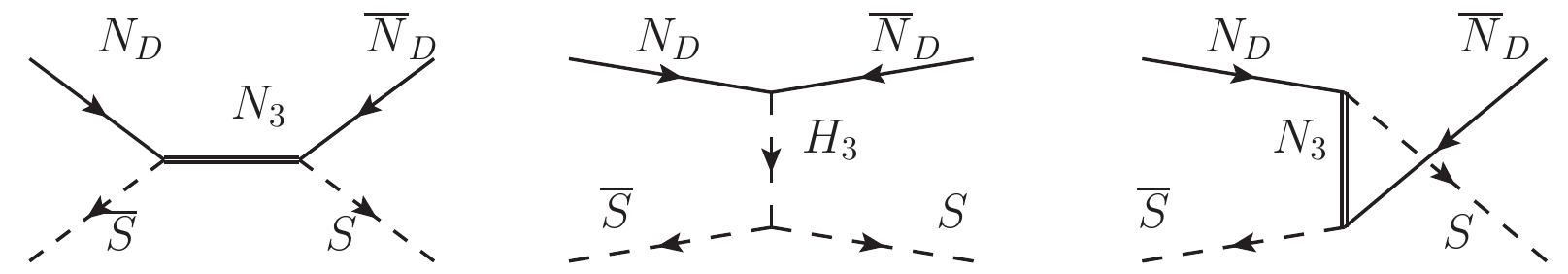}\,\\
{\rm $b)$}\includegraphics[width=0.8\textwidth]{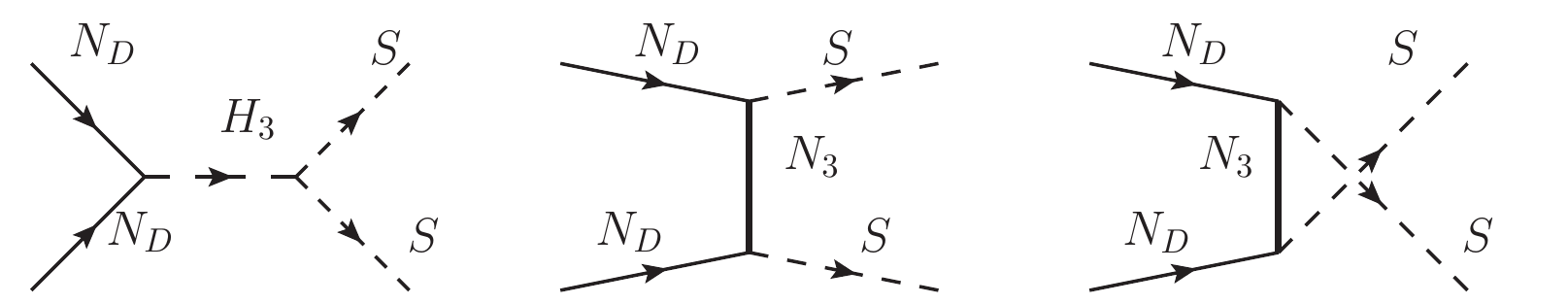}\,
\caption{Feynman diagrams of the $\Delta N_{D}=\D S=2$ scatterings.}
\label{DN2scatt}
\end{center}
\end{figure}

\begin{figure}[h!]
\begin{center}
{\rm $a)$}\includegraphics[width=0.45\textwidth]{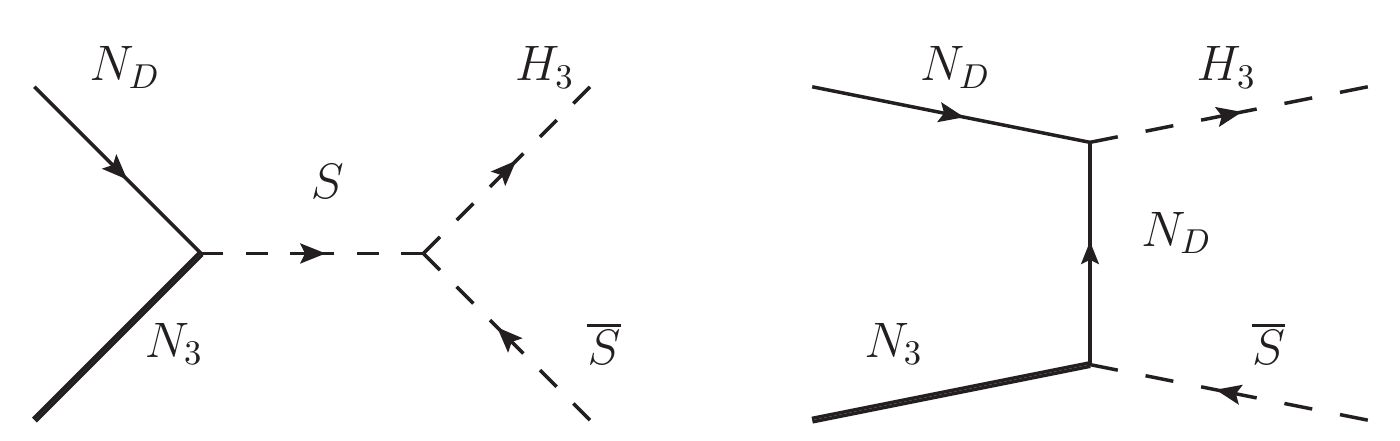}\,,\hspace{2mm} {\rm $b)$} \includegraphics[width=0.45\textwidth]{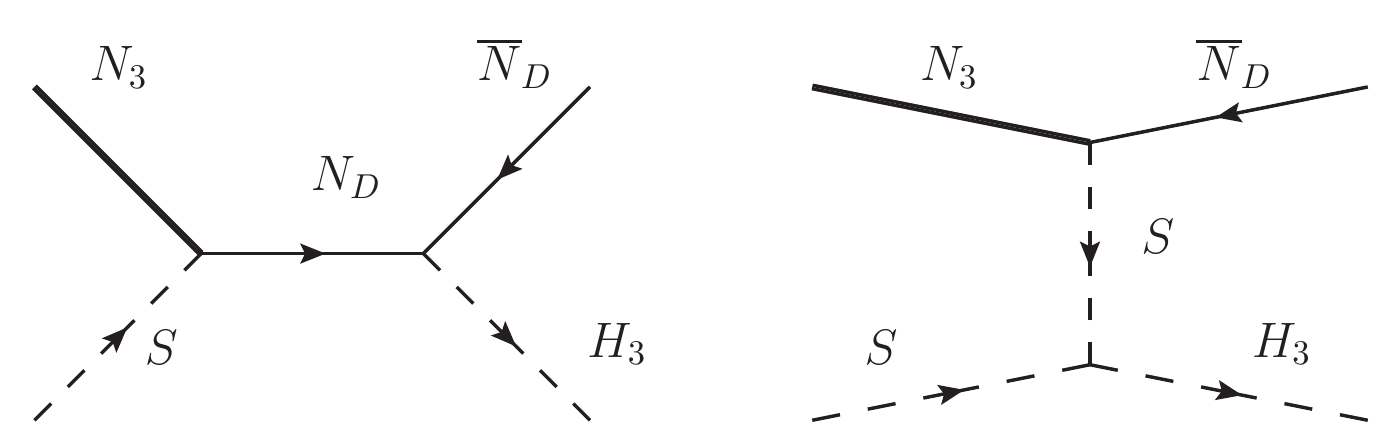}\\
{$c)$}\includegraphics[width=0.5\textwidth]{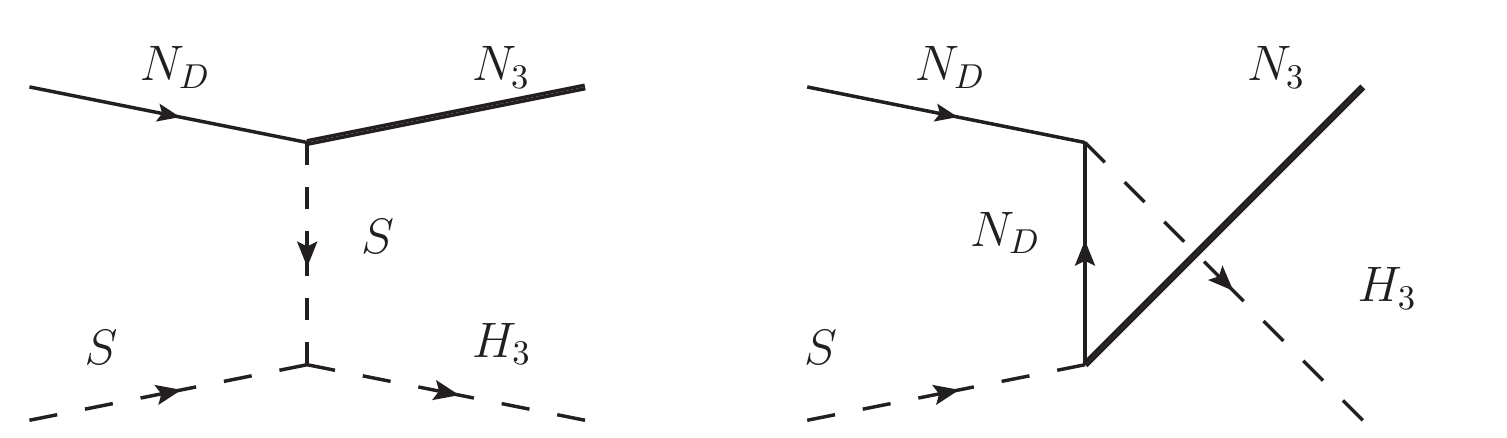}
\caption{Feynman diagrams of the $\Delta N_{D}=\D S=1$ scatterings.}
\label{DN1scatt}
\end{center}
\end{figure}

\begin{figure}[h!]
\begin{center}
\includegraphics[width=0.45\textwidth]{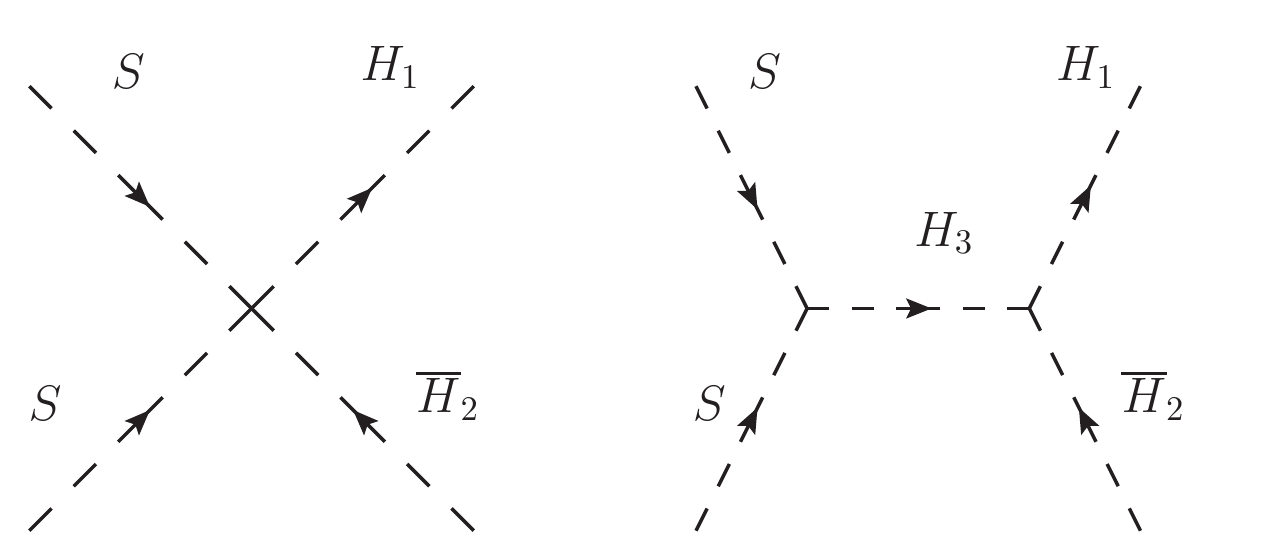}
\caption{Feynman diagram of the $S$ self-annihilation.}
\label{SSelfAnn}
\end{center}
\end{figure}

\subsubsection{First Stage: processes at ${\mathcal O}(\al^2)$, ${\mathcal O}(g^2)$, ${\mathcal O}(g^2\,\al^{2})$ and  ${\mathcal O}(g^4)$. }

We list below the processes relevant in the first step,  where the asymmetries in $S$ and $N_D$ are created.~\footnote{We denote by $\Delta X$ the absolute variation of the $X$ particle number density.}~Further details are given in Appendix \ref{BESec}:
\begin{itemize}
\item Decays and inverse decays of $N_{3}$: $N_3 \to N_D\,\ovl{S}\,,\ovl{N}_{D}\,S$ (see Fig.~\ref{CPAsymmGraph}). 
\item $\Delta N_D=\Delta S=2$ scatterings: $N_D\,\ovl{S}\leftrightarrow \ovl{N}_D\,S$ and $N_D\,N_D \leftrightarrow S\,S$ (see Fig.~\ref{DN2scatt}).
\item $\Delta N_D=\Delta S=1$ scatterings:  $N_D\,N_3\leftrightarrow H_3\,\ovl{S}$, $N_D\,\ovl{H}_3\leftrightarrow N_3\,\ovl{S}$ and  
$N_D\,S\leftrightarrow N_3\,H_3$ (see Fig.~\ref{DN1scatt}). 
\item $S$ self-annihilation: $S\,S\leftrightarrow H_{1}\,\ovl{H}_{2}$ (see Fig.~\ref{SSelfAnn}).
\end{itemize}
Notice that the last process depends on interaction terms reported in the 
scalar potential of the model (see eq.~(\ref{VDM})). However, it turns out to be numerically irrelevant, so we disregard the effect of this term in the following.

\begin{figure}[t!]
\begin{center}

\includegraphics[width=1.\textwidth]{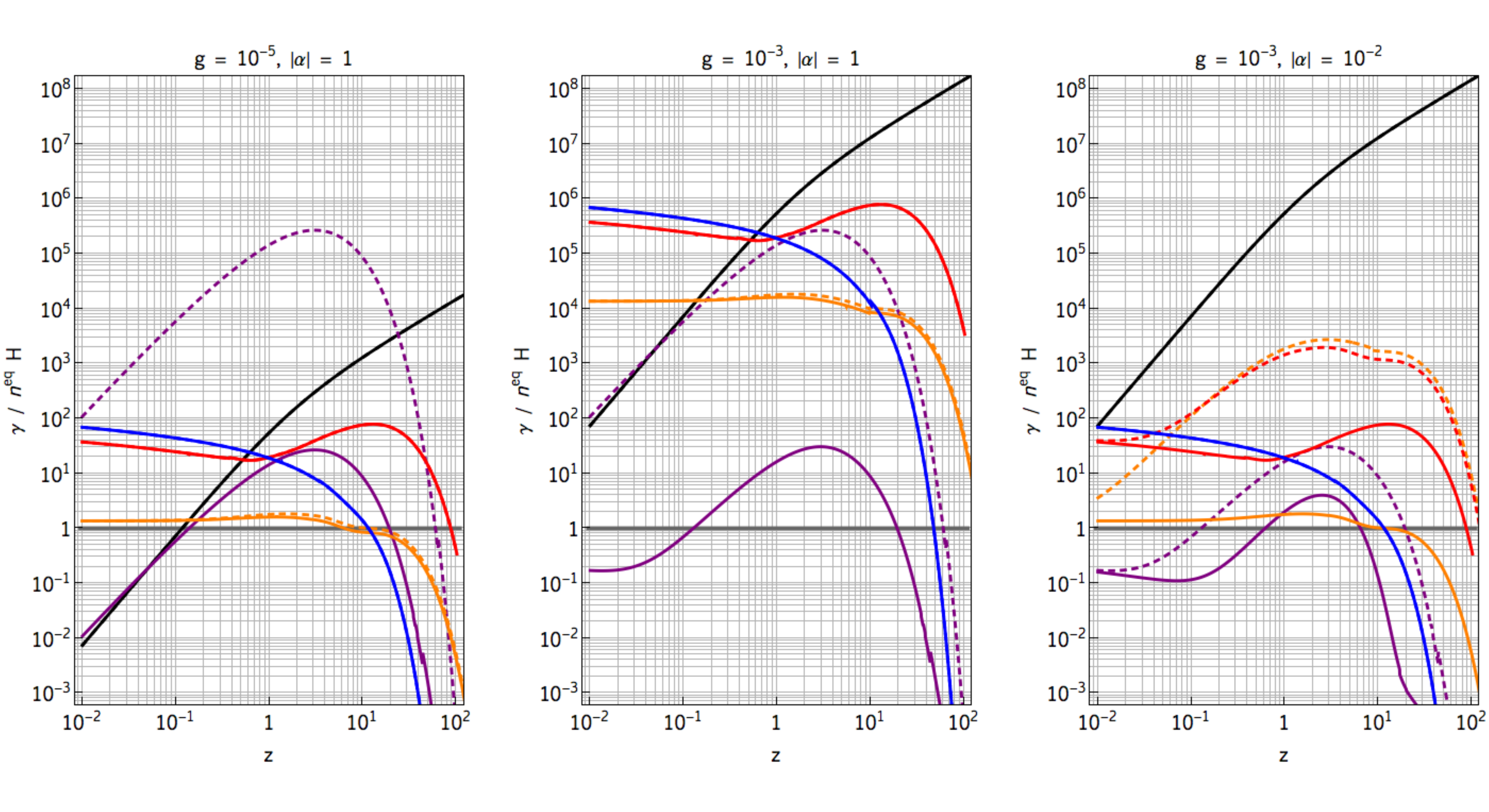}
\caption{Processes relevant in the first stage of leptogenesis: thermal density rates as function of $z\equiv M_{3}/T$,
for $M_3=50$ TeV, $M=10$ TeV and $\mupp = 1 (100)$ GeV, plain (dashed) curves. In black is reported the total decay
rate of $N_{3}$. The purple curves stand for the sum of the (non-resonant part of the) diagram $a)$ and diagram $b)$ in Fig.~\ref{DN2scatt}.
The blue, orange and red curves correspond, respectively, to the processes $a)$, $b)$ and $c)$ shown in Fig.~\ref{DN1scatt}.}\label{GrapheCS1}
\end{center}
\end{figure}

We display in Fig.~\ref{GrapheCS1} the interaction rates $\gamma^{eq}$ of some of the up-listed processes 
as function of the parameter $z\equiv M_{3}/T$, where $T$ is the temperature of the 
plasma. These rates are normalized by $H(z)\,n_{N_{3}}^{eq}(z)$, except for the $\Delta N_{D}=2$ rates which are normalized by $H(z)\,n_{N_{D}}^{eq}(z)$, as they only act as damping terms.~\footnote{In Fig.~\ref{GrapheCS1} only the off-shell part of the $\Delta N_{D}=2$ diagrams $a)$ is shown, as its on-shell part equals $\gamma_{D}/4$, $\gamma_{D}$ being the total decay rate of $N_{3}$. }
For illustration we fix $M_3=50\TeV$, $M=10 \TeV$ and we choose representative values of $g$ and $\vert \al \vert$ for the different panels.~\footnote{For definiteness, in the following numerical evaluations we set the phase of $\alpha$ to its maximum value $\alpha = -\rm{i}\,\vert \alpha \vert$.}  

We represent in each plot by straight (dashed) lines the computed rates assuming $\mupp=1\, (100)\GeV$. For $\mupp=1 \GeV$, the cross-sections of the $\D N_{D}=2$ scatterings $a)$ are dominated by their $s$- and $u$-channels and scale as $\mathcal{O}(g^4)$. The $\D N_{D}=2$ scatterings $b)$ on the other hand are governed by their $s$-channel and are proportional to $\vert\al \vert^{2}\,\mu^{\prime \prime\,2}$. The $\D N_{D}=1$ processes $a)$, $b)$ and $c)$ of Fig.~\ref{DN1scatt} are dominated by their respective $t$-, $s$- and $u$-channels and therefore scale as $\mathcal{O}(g^{2}\,\vert \al \vert^2)$. 
For larger values of $\mupp$, \textit{e.g.} $\mupp=100\GeV$, the $\Delta N_{D}=1$ processes $b)$ and $c)$ get sizable contributions from  their $t$-channels ($\propto g^2\,\mu^{\prime \prime\,2}$) which dominate over the other channels for small values of $\alpha$, as can be seen in the right panel of Fig.~\ref{GrapheCS1}.
The different interaction rates where evaluated using the packages FeynArts \cite{FeynArts} and FormCalc \cite{FormCalc}. To this end, we implemented our model, $\mathcal{L}_{\rm int}$ eq.~(\ref{Linter}), via FeynRules~\cite{FeynRules}.

The various interactions considered above control the amount of $N_{D}$ and $S$ asymmetries produced during the first stage of leptogenesis. 
As the lepton asymmetry -and finally the baryon asymmetry- mostly depends on the amount of $N_{D}$ asymmetry produced in the first step, it is useful to introduce an efficiency factor $\eta_{1}$ defined through:
\bea
\label{DefineEta1}
Y_{\D N_{D}}(z_{tr}) =\eps_{CP}\,\eta_{1}\,Y_{N_3}^{eq}(T\gg M_3)\,.
\eea
In this parametrization, $Y_{X}$ indicates the comoving number density of $X$, while $z_{tr} \sim M_3/M$ approximately marks the transition between the first and second stage: for $z\gtrsim z_{tr}$, \ie \, $T\lesssim M$, $N_D$ decouples from the plasma and decays into leptons and antileptons.  

\begin{figure}[t!]
\begin{center}
\includegraphics[width=0.5\textwidth]{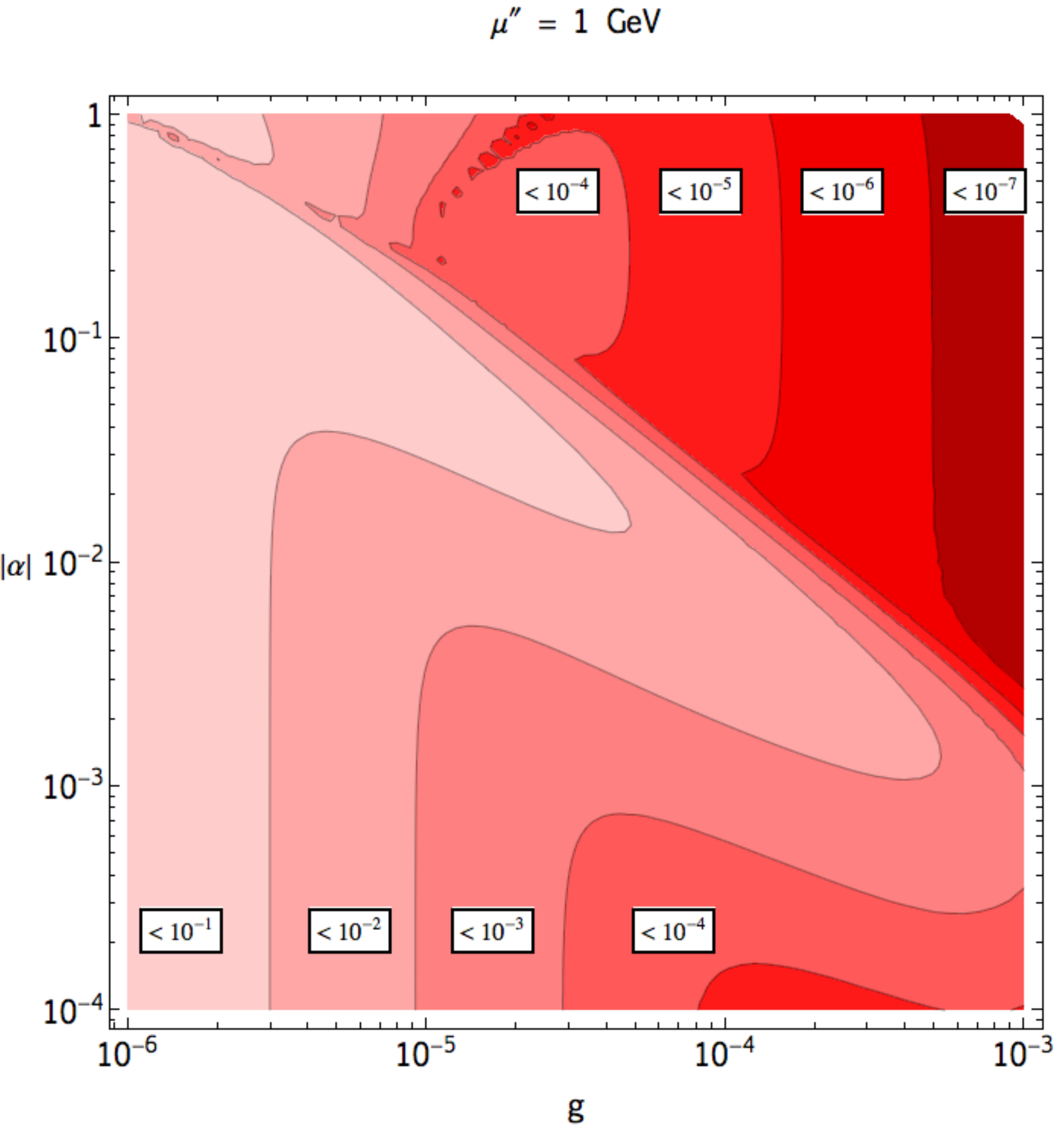}\,\includegraphics[width=0.5\textwidth]{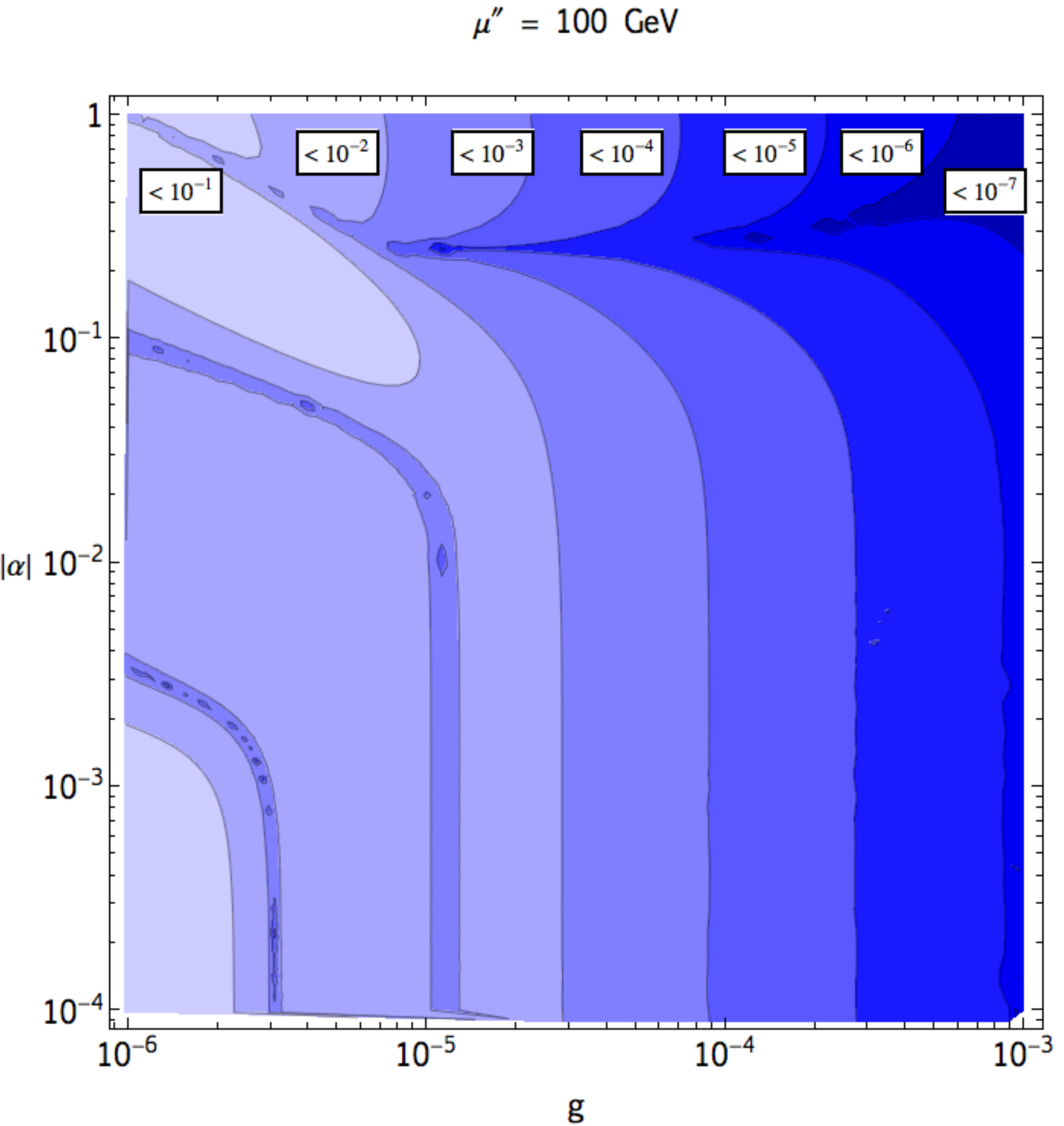}
\caption{Efficiency $\eta_{1}$ of production of $N_{D}$ asymmetry in the first stage of leptogenesis, cf. eq.~(\ref{DefineEta1}), as a function of $g$ and $\vert \al \vert$ for  
$M_3=50\TeV$, $M=10 \TeV$ and $\mupp=1 (100)\GeV$ in the left (right) panel. }
\label{GrapheEff1}
\end{center}
\end{figure}

Given the numerous interactions considered above, the derivation of an analytic expression for the efficiency factor $\eta_{1}$ is quite challenging.
Nevertheless, we perform a numerical evaluation of $\eta_{1}$ by solving the set of Boltzmann equations reported in Appendix~\ref{BESec}. The resulting efficiency is shown in Fig.~\ref{GrapheEff1}, where iso-contours of $\eta_{1}$ in the $g-\vert\alpha\vert$ plane are displayed, for $M_3=50\TeV$, $M=10 \TeV$ and $\mupp= 1 \GeV$  ($100 \GeV$) in the left (right) panel.

We first consider the case of small $\mupp$, left panel of Fig.~\ref{GrapheEff1}. In this case, the $\D N_{D}=2$ scatterings are typically smaller than the decays and inverse decays, as shown in Fig.~\ref{GrapheCS1}. Depending on the value of $\alpha$, the $\Delta N_{D}=1$ scattering rates $\gamma_{N_3}^{k}$ $(k=a,b,c)$, may be in equilibrium when the $N_{D}$ asymmetry is produced. This occurs if 
\bea
\frac{\gamma_{N_3}^{k}}{n_{N_D}^{eq}\,H(M_3)}\gtrsim 1 \,\quad \Longrightarrow \,\quad \vert \al \vert \times \left(\frac{g}{10^{-6} }\right)\gtrsim 1\,.\label{cond1}
\eea
As can be seen in the left panel of Fig.~\ref{GrapheEff1}, the efficiency $\eta_{1}$ strongly depends on whether the $\D N_{D}$ scatterings are in equilibrium at $T\sim M_{3}$ or not. 
In the case their rates are not fast enough, \ie \,if the condition~(\ref{cond1}) is not satisfied, the production of $N_D$ and $S$ asymmetries is mostly driven by decays and inverse decays of $N_{3}$.
This situation is very similar to the standard leptogenesis scenario, when $\D L=1$ scatterings are neglected.
Therefore, we expect that larger values of the coupling $g$ increase the washout effects. The strength of $N_{3}$ decays and inverse decays can be expressed in terms of the washout parameter $K_{D}$
\bea
K_D\equiv \frac{\Gamma_{N_3}}{H(M_3)} \simeq 2\,\left(\frac{g}{10^{-6}}\right)^2 \left(\frac{50 \TeV}{M_3}\right)\,.
\eea
For $g\gtrsim 10^{-6}$ and $M_{3}\sim \mathcal{O}(10)\TeV$, decays and inverse decays act in a strong washout regime, where the efficiency  is approximately given by~\cite{Buchmuller:2004nz}:
\bea
\eta_{1}\sim  \frac{0.4}{\,K_D\,\log(K_D) }\,.\label{eff1}
\eea 
For smaller values of $g$, decays and inverse decays act in a weak washout regime, and the efficiency scales as $K_{D}^2$~\cite{Buchmuller:2004nz}, in the case where the abundance of $N_{3}$ is vanishing at high temperatures.

In the opposite regime, when the $\Delta N_{D}=1$ scatterings are fast enough and the condition eq.~(\ref{cond1})  is satisfied, an initial (anti-)asymmetry is produced at earlier times, due to the $CP$ violation in scatterings, which is discussed in Appendix~\ref{BESec}. From Fig.~\ref{GrapheEff1}, we can distinguish two relevant cases, according to the values of $\alpha$ and $g$. 
For $\vert \al \vert  \approx 10^{-6}/g $, the $\D N_{D}=1$ scatterings essentially act as source terms, producing $N_{D}$ and $S$ asymmetries, thereby increasing the efficiency $\eta_{1}$. This effect is manifest in the diagonal of the left plot of Fig.~\ref{GrapheEff1}. Conversely, for larger values of $\al$, the $\D N_{D}=1$ scatterings act as damping terms and increase the washout of the asymmetries. The resulting efficiency is therefore highly reduced.

The case of a larger $\mupp$ is depicted in the right panel of Fig.~\ref{GrapheEff1}, where we fix $\mupp=100 \GeV$, while $M_{3}$ and $M$ assume the same values as before.
As already stated, in this case the $\D N_{D}=1$ scatterings $b)$ and $c)$ pick-up  sizable contributions from their corresponding $t$-channels, and are enhanced for relatively small values of $\alpha$, compared to the $\mupp=1$ GeV case, as can be seen in the right panel of Fig.~\ref{GrapheCS1}.
For $\vert \al \vert \lesssim 0.1$, the efficiency depends essentially on $g$, since the scatterings are negligible with respect to the decays and inverse decays, and then $\eta_{1}$ behaves as in eq.~(\ref{eff1}). However, for small values of $g$, $g\lesssim$ few$\,10^{-5}$, $\D N_{D}=1$ scatterings become competitive with decays and inverse decays, both in the generation and in the washout of $N_{D}$ and $S$ asymmetries. For values of $\alpha$ of order one and small $g$, the scatterings mainly act as source terms in analogy with the $\mupp=1$ GeV regime, thus increasing $\eta_{1}$. 

We see that in this first stage, the efficiency of $N_{D}$ and $S$ asymmetry production can be close to its maximum possible value in a large region of the parameter-space. However, this does not guarantee a successful leptogenesis, as this asymmetry should be transferred efficiently to leptons.

\subsubsection{Second Stage: processes at ${\mathcal O}(y_{1,2}^2)$, ${\mathcal O}(g^2\,y_{1,2}^{2})$.}

We now concentrate on the second step of leptogenesis: the transfer of $N_D$ asymmetry to the lepton doublets.
Once a lepton asymmetry is generated, the sphaleron processes which are active at the leptogenesis epoch convert part of it into a non-zero baryon number density. The second stage ends at the freeze-out of the sphalerons, that may occur before or right after EWSB~\cite{Harvey}.

We report below the main $\D \ell =1$ processes which participate in the lepton charge transfer mechanism:
\begin{itemize}

\item Decays of $N_D$, which are either $L$-conserving,  $N_D\to \ell\, H_1$, or $L$-violating, $N_D \to \ovl{\ell}\,\ovl{H}_2$. 

\item Scatterings on top quarks: the $s$-channel $N_{D}\,\ovl{\ell} \leftrightarrow \ovl{t}\,q_{3}$ and the $t$-channels $N_{D}\,q_{3}\,(\ovl{t})\leftrightarrow \ell\, t \,(\ovl{q_{3} })$. These processes are mediated by the exchange of the Higgs doublet $H_1$ and correspond to the $\D L=1$ scatterings in standard leptogenesis. Notice, however, that in our case lepton number is conserved.

\item Scatterings on $N_3$: $N_3\,S\leftrightarrow \ell\,H_1$ and $N_3\,S\leftrightarrow \ovl{\ell}\,\ovl{H}_2$ which are mediated by $N_D$. A $CP$ asymmetry emerges from these processes, as shown in Appendix~\ref{BESec}.
\end{itemize} 
We do not include  the scatterings involving gauge bosons in our evaluation of the baryon asymmetry.  However, we do not expect these processes to have a quantitative impact. Indeed, they cannot act as a source term for the lepton asymmetry since no $CP$ violation is possible in this case, in contrast to the standard leptogenesis scenario~\cite{GaugeCP}. In addition, they tend to equilibrate the lepton and $N_{D}$ number densities, like the scatterings on top quarks considered above. Actually, it is shown in references \cite{GiudiceLepto}-\cite{Pilaftsis05} that these processes have comparable rates.

The lepton doublet can also participate in $\D L=2$ $N_{D}$-mediated scatterings, similarly to  the standard leptogenesis case: $\ell\, H_1 \leftrightarrow \ovl{\ell}\,\ovl{H}_{2}$ and $\ell\,\ell \leftrightarrow \ovl{H}_{1}\, \ovl{H}_{2}$. In this case the scattering rate is proportional to both the neutrino Yukawa couplings, $y_{1}$ and $y_{2}$.  In a democratic scenario, that is for $|y_{1}|\approx |y_{2}|$, provided the constraints from active neutrino masses, eqs~(\ref{nueig})
and (\ref{eta12cond}) are satisfied, such $\Delta L=2$ scatterings are usually in equilibrium at the leptogenesis time. They are however greatly suppressed compared to the $\Delta \ell=1$ scatterings and turn out to be numerically irrelevant, as illustrated below.

In this second stage, all interactions depend on the neutrino Yukawa couplings $y_1$ and $y_2$. 
In the limit where these couplings are zero, no lepton (doublet) asymmetry can be generated as basically both $N_D$ and $S$ decouple from the SM sector.
This clearly implies a lower bound on the values of $y_1$ and $y_2$.

\begin{figure}[t!]
\begin{center}
\includegraphics[width=0.6\textwidth]{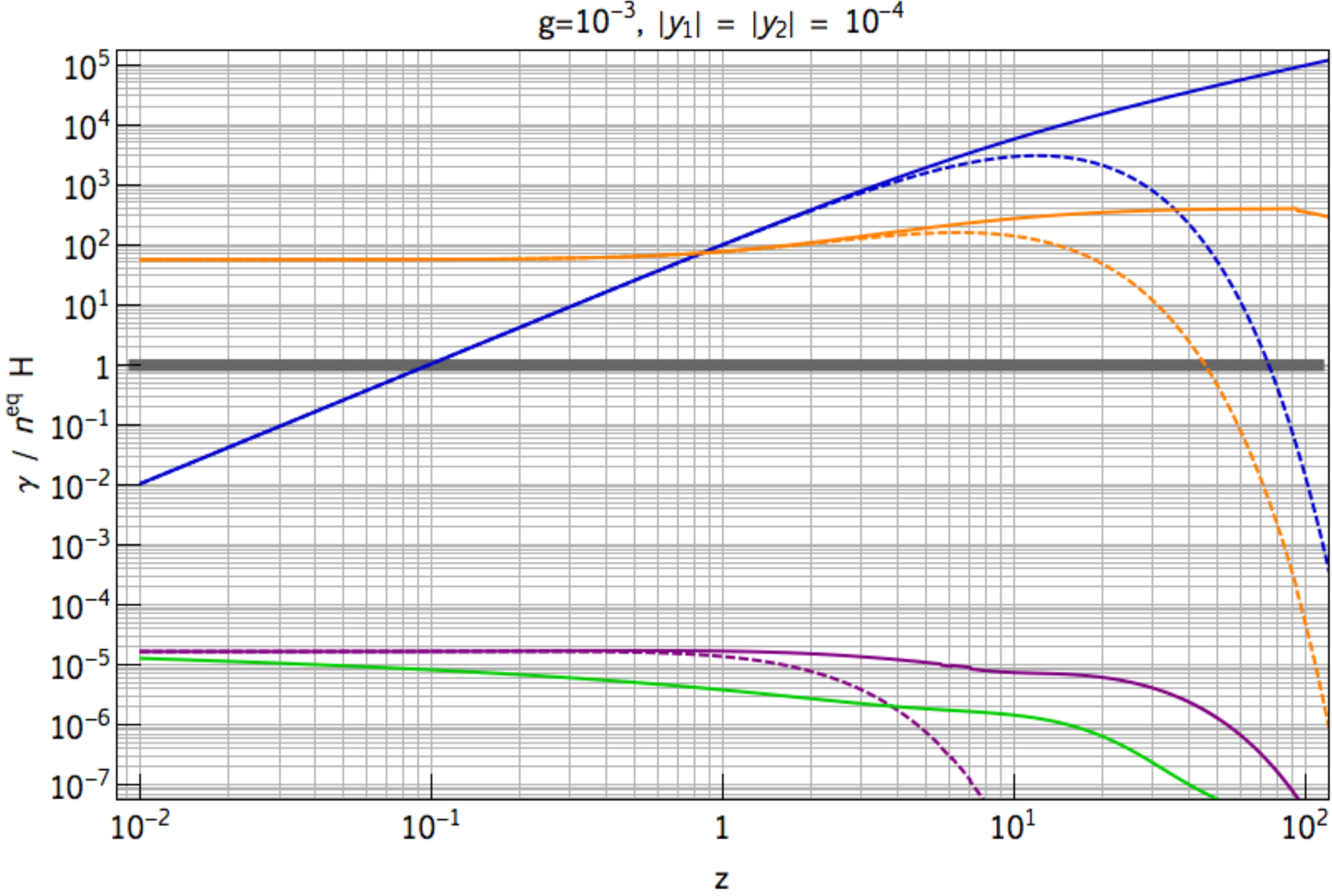}
\caption{Processes relevant in the second stage of leptogenesis: thermal density rates $\gamma^{eq}$ as function of $z\equiv M_{3}/T$,
for $M_3=50\TeV$, $M=10 \TeV$. The blue curves stand for decays of $N_{D}$. The orange curves stand for the (sum of $s$- and $t$-channel) $\Delta \ell=1$ scatterings on top quarks, while the purple ones represent $\Delta \ell=1$ scatterings on  $N_3$. The green line corresponds to the total $\D L=2$ scattering rate.}\label{GrapheCS2}
\end{center}
\end{figure}

Let us discuss this bound, independently of the constraints from low-energy neutrino masses, as it sheds light on how this second stage works.
To this end, we represent in Fig.~\ref{GrapheCS2} the processes relevant in the second step for the same set of parameters used in Fig.~\ref{GrapheCS1}:
$M_3=50 \TeV$, $M=10 \TeV$ and $g=10^{-3}$, and for $|y_1| =|y_{2}|=10^{-4}$.
We represent in Fig. \ref{GrapheCS2} by plain (dashed) curves the rates normalized by $n_{X}^{eq}\,H(M)$, where $X= N_{D}~ (\ell)$, acting as source (damping) terms.
The blue curves correspond to $N_{D}$ decays, both $L$-conserving and violating as $\vert y_{1} \vert=\vert y_{2}\vert$. The orange (purple) curves are related to scatterings on top ($N_{3}$), while the green line stands for the $\Delta L=2$ processes. Scatterings on $N_{3}$ and $\Delta L=2$ interactions are clearly sub-dominant and can be neglected.

A lower bound on $y_1$ can be derived  by demanding that the scattering rates on top quarks, denoted by $\gamma_{N_{D}}^{t}$, are in equilibrium  at $T\sim M$, when acting as a source term for the lepton asymmetry:
\bea\label{CondScat}
\frac{\gamma_{N_D}^{t}}{n_{N_D}^{eq}\,H(M)} \gtrsim 1 \quad \Longrightarrow \quad \vert y_1 \vert\gtrsim 10^{-5}\times\sqrt{\frac{M}{10\,\TeV}}\,.\label{cond2}
\eea
Therefore, provided $y_1$ is large enough, the $L$-conserving scatterings are in equilibrium and can transfer the $N_{D}$ asymmetry to the lepton doublets.
A similar lower bound arises for $y_{2}$ from the corresponding $\Delta L=1$ scatterings with gauge bosons.

The main source of lepton asymmetry production may originate just from the decays of $N_D$.
Let us suppose, indeed, that the lower bound on $y_1$, eq.~(\ref{CondScat}), is not satisfied. Still, as we see from Fig. \ref{GrapheCS2}, decays dominate over the scatterings at $T\sim M$.
For these decays to be effective in redistributing the $N_{D}$ asymmetry to leptons, the Dirac neutrino should be heavy enough, say $M\gtrsim 10\times T_{sph}$, and the following condition should be satisfied:
\bea \label{cond3}
\Gamma_{N_D} \gtrsim H(M) \quad \Longrightarrow \quad \vert y_{1,2} \vert \gtrsim 6\times 10^{-7}\,\sqrt{\frac{M}{10\TeV}}\,.
\eea

In summary, for neutrino Yukawa couplings smaller than the bound above, the lepton number asymmetry production is not efficient. if only condition (\ref{cond3}) is satisfied, almost all $N_{D}$ decays to leptons, and we expect that at the end of the second stage, the lepton asymmetry equals the amount of $N_{D}$ asymmetry produced in the first stage. For larger Yukawa couplings satisfying eq. (\ref{cond2}), $Y_{\D \ell }(z_{tr}) \approx Y_{\D N_{D}}(z_{tr})$ at the end of the first stage, so at the end of the second stage, $Y_{\D \ell}(z_{sph}) \approx 2\,Y_{\D N_{D}}(z_{tr})$. 

\subsubsection*{The case of a light Dirac Neutrino}
\begin{figure}[t!]
\begin{center}
\includegraphics[width=0.7\textwidth]{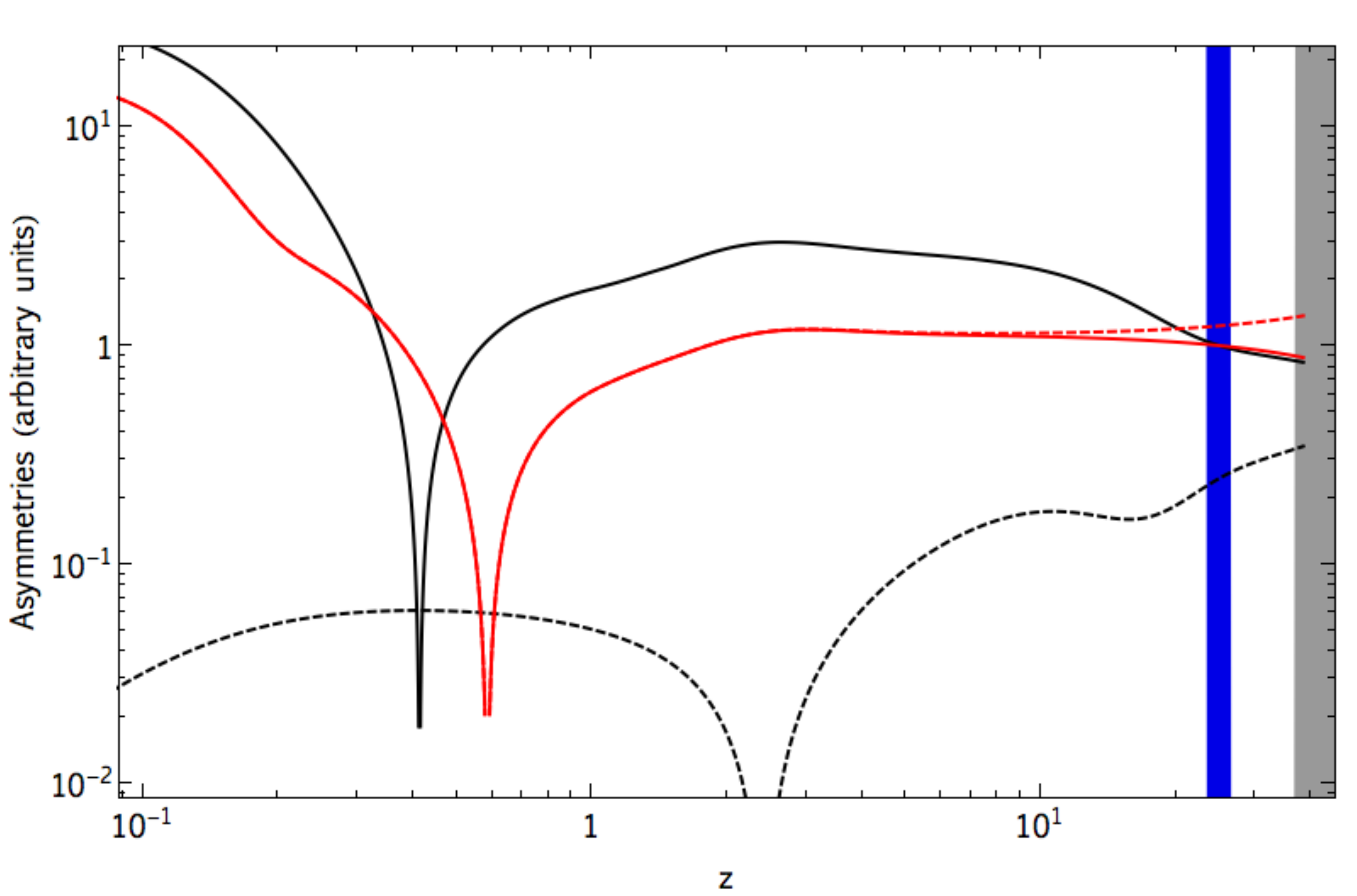}
\caption{Influence of scatterings on the transfer of $N_D$ asymmetry to a lepton doublet asymmetry. See the text for details.}\label{GrapheSCA}
\end{center}
\end{figure}

An interesting case is realized when the Dirac neutrinos are so light that they don't have enough time to decay before the freeze-out of the sphalerons.
Demanding that the scatterings with quarks are in equilibrium, at least slightly before the sphalerons decouple, the condition (\ref{cond2}) is changed to $\vert y_1 \vert\gtrsim 10^{-6}$.
Therefore $y_{1}$ should be at least larger than the electron Yukawa coupling. We illustrate this remarkable case in Fig.~\ref{GrapheSCA}, where we fix $M=200 \GeV$ and we consider two sets of values for $|y_{1}|$ and $|y_{2}|$: $i)$ $|y_{1}|=|y_{2}|=5\times10^{-4}$ (red curves) and $ii)$ $|y_{1}|=|y_{2}|=5\times 10^{-7}$ (black curves). 
Notice that the latter case may hardly be compatible with constraints from neutrino masses, eqs~(\ref{nueig})-(\ref{v2low}).
The dashed and plain curves correspond to the lepton doublet and $N_D$ asymmetries, respectively, and all the asymmetries have been normalized to $Y_{N_{D}}(z_{tr})$,
where $z_{tr}\sim M_{3}/M$ is indicated by a blue band. We impose a sphaleron freeze-out at around $T_{sph}\sim 130\GeV$, which is represented by the gray band. \\
We see from Fig.~\ref{GrapheSCA} that while $N_D$ asymmetry is almost unaffected by the Yukawa hierarchy, in case $ii)$ leptons do not equilibrate with $N_{D}$ 
as scatterings are out-of-equilibrium, while in case $i)$ $Y_{\D \ell} \approx Y_{\D N_D}$ at temperatures well above $M$.

Provided that the neutrino Yukawa couplings are sufficiently large, an asymmetry in $N_{D}$ will be always transmitted to the lepton sector, regardless of the Dirac neutrino mass: we can therefore asset that no lower bound can be derived on $M$ from leptogenesis.\\

In conclusion, once light neutrino mass constraints are applied, a lepton asymmetry is efficiently produced. A successful leptogenesis then only relies upon the  amount of $N_{D}$ asymmetry produced in the first stage.

\subsection{Successful leptogenesis}
\begin{figure}[t!]
\begin{center}
\includegraphics[width=0.5\textwidth]{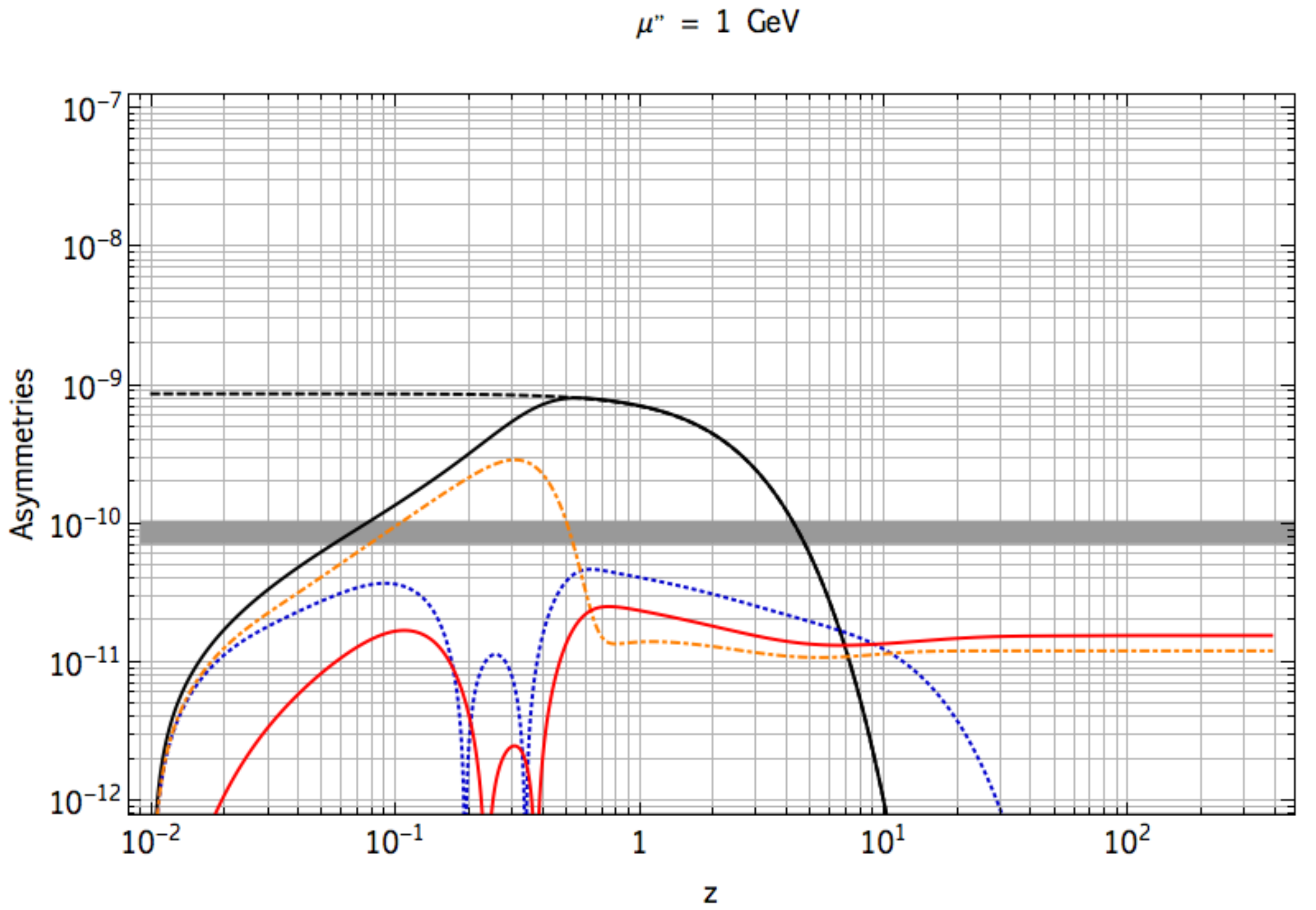}\,\includegraphics[width=0.5\textwidth]{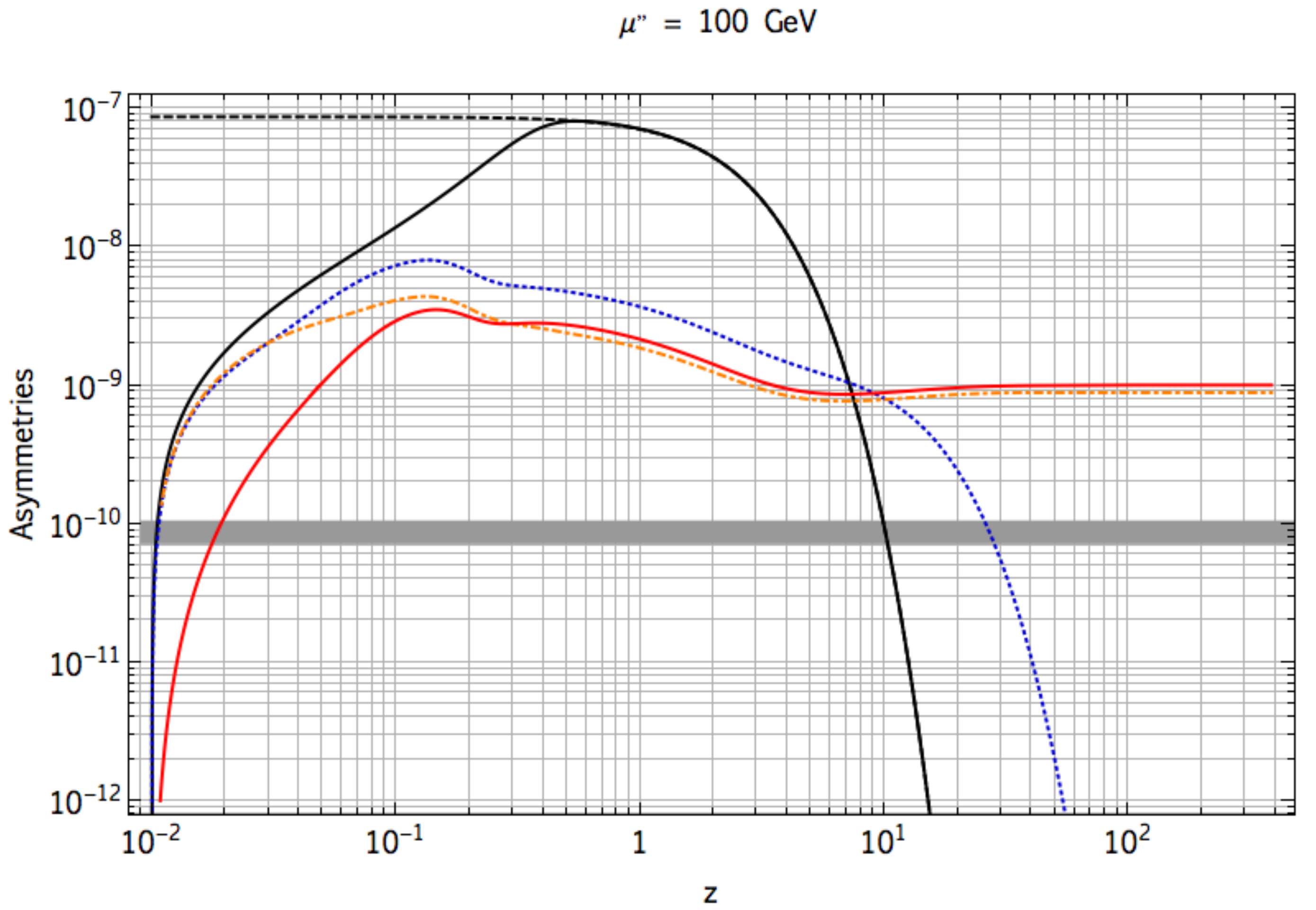}
\caption{Evolution of number density asymmetries in function of $z$. $M_{3}=50$ TeV, $M=10$ TeV, $\vert \al \vert =0.1$ and $g=10^{-5}$ are fixed. The black plain (dashed) curves represent $ \eps_{CP} \times Y_{N_{3}}^{(eq)}$. In dotted-blue, dot-dashed orange and plain red are shown $Y_{\Delta N_{D}}$, $Y_{\Delta S}$ and $Y_{\Delta B}$ respectively.}
\label{GrapheYB}
\end{center}
\end{figure}
In the former subsections, we analyzed the conditions under which a $N_{D}$ asymmetry is efficiently produced during the first step of leptogenesis, and subsequently transmitted to the lepton doublets. Through the sphaleron processes, this lepton asymmetry is partly converted into a baryon number density. The sphalerons violate both lepton and baryon numbers, but conserve $B-L$: it is therefore more convenient to evaluate the $B-L$ asymmetry. Given the different processes in thermal equilibrium during leptogenesis era, the final baryon asymmetry reads:
\bea\label{YB}
Y_{\D B}=\frac{2}{7}\,Y_{\D (B-L)}(z_{sph})\,.
\eea
The derivation of 
eq.~(\ref{YB})  is given in Appendix~\ref{ChemEqCon}.
\begin{figure}[t!]
\begin{center}
\includegraphics[width=0.5\textwidth]{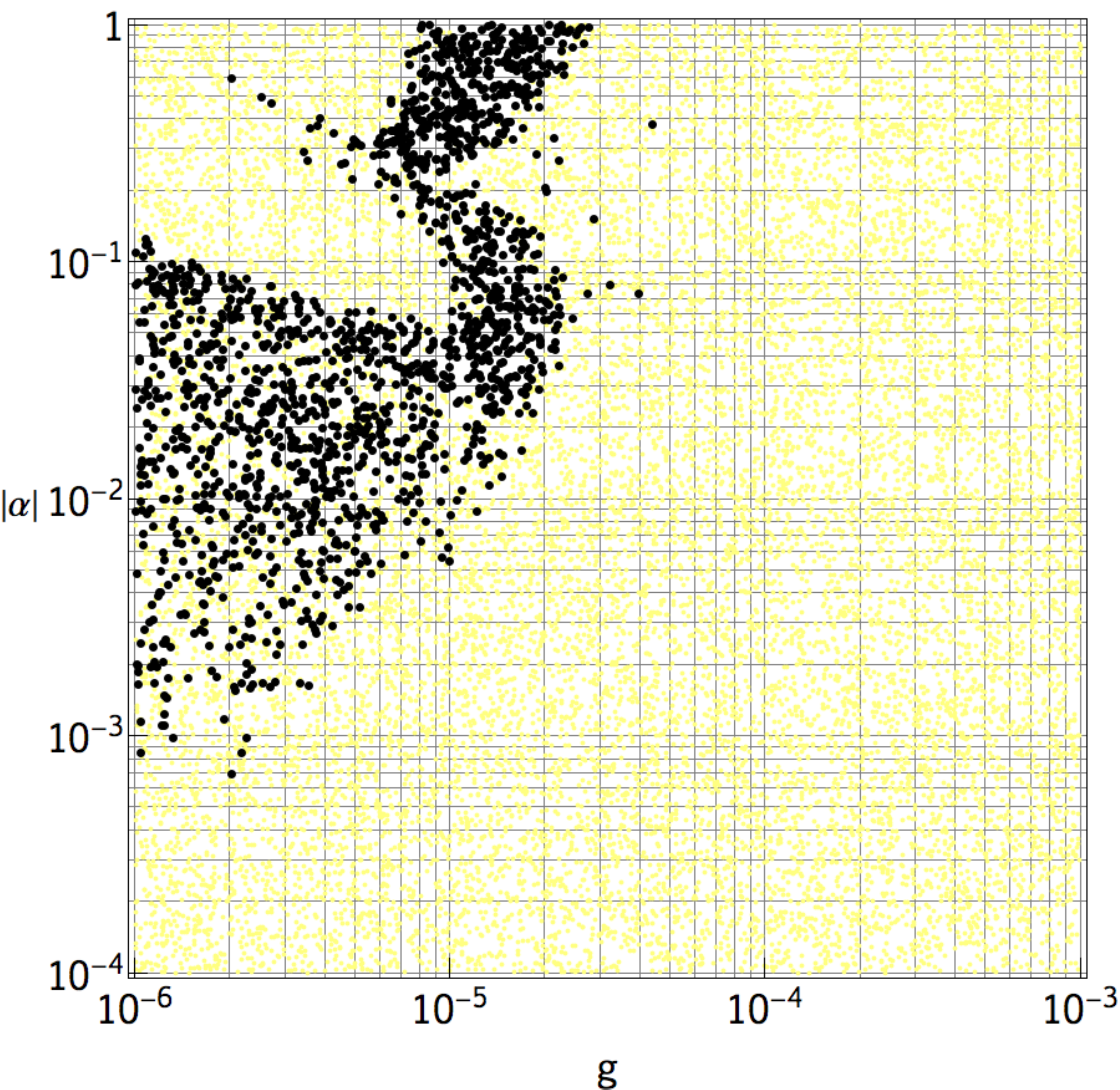}
\caption{Successful leptogenesis. Region of the parameter-space in the $\vert \al \vert$-$g$ plane providing a final baryon asymmetry (not) compatible observations, (yellow) black points. We fix $M_{3}=50$ TeV, $M=10$ TeV, $\mu_{S}=100$ GeV and $\mupp=100$ GeV.}
\label{GrapheAG2}
\end{center}
\end{figure}
We illustrate in Fig.~\ref{GrapheYB} the evolution of $N_{D}$, $S$ and baryon asymmetries against $z$ for typical values of the parameters: $M_{3}=50$ TeV, $M=10$ TeV, $\mu_{S}=100$ GeV and $\mupp= 1$ or $100$ GeV, left or right panel respectively, for fixed values $\vert \alpha \vert =0.1$ and $g=10^{-5}$. 
For such values of $\vert \al \vert$ and $g$, we see from Fig.~\ref{GrapheEff1} that the efficiencies $\eta_{1}$ are quite similar, that is $\eta_{1}\sim 0.1$. However, for $\mupp=100$ GeV the $CP$ asymmetry is $\eps_{CP}\simeq 4\times 10^{-6}$, two orders of magnitude larger than for $\mupp=1$ GeV, cf. eq.~(\ref{ApprECP}), and so the baryon asymmetry in the former case will be bigger. Indeed, for $\mupp=1$ GeV, $Y_{\Delta B}\approx 1.6\times 10^{-11}$ while for $\mupp=100 \GeV$, $Y_{\Delta B}\approx 10^{-9}$. These values should be compared with the measurement of WMAP~\cite{WMAP}:
\bea
Y_{\Delta B}^{\rm obs}=\left(8.77\pm 0.21\right)\times10^{-11}\,.
\eea
In Fig.~\ref{GrapheAG2}, we made a scan over the two parameters $\alpha$ and $g$: the black points represent values of $Y_{\Delta B}$ compatible with observations.~\footnote{Actually, for the sake of illustration, we enlarge the required range, demanding $3\times 10^{-11} \lesssim Y_{\Delta B} \lesssim 3\times 10^{-10}$.}
As we see, a successful leptogenesis is easily realized  in our scenario, provided the $CP$ asymmetry is big enough, that is $\eps_{CP} \gtrsim 3\times 10^{-7}$, and the washout processes do not suppress the $N_{D}$ asymmetry in the first stage,  $i.e.$ $\eta_{1} \gtrsim$ few $10^{-3}$.

\section{The Scalar Sector}\label{ScalarSect}

Given the charge assignment of the scalar fields in Tab.~\ref{FieldAssign}, the most general scalar potential $\mathcal{V}_{\rm SC}$ invariant under $SU(2)_{W}\times U(1)_{Y}\times \left[U(1)_{B-\tilde{L}}\right]$ can be written in the following form
\bea
	\mathcal{V}_{\rm SC}&\equiv& \mathcal{V}_{\rm SB}\,+\,\mathcal{V}_{\rm DM}\,,
\eea
where $\mathcal{V}_{\rm SB}$ and $\mathcal{V}_{\rm DM}$ denote the symmetry breaking and dark matter scalar potentials, respectively:
\bea
\mathcal{V}_{\rm SB}&=&-\mu_{1}^2\, H_1^{\dag}\,H_1 + \lam_{1}\, (H_1^{\dag}\,H_1)^2 - \mu_{2}^2\, H_2^{\dag}\,H_2 + \lam_{2}\, (H_2^{\dag}\,H_2)^2 - \mu_{3}^2\, H_3^{*} H_3 + \lam_{3}\, (H_3^{*} H_3)^2\nonumber \\  &+& \kappa_{12}\, H_1^{\dag}\,H_1 H_2^{\dag}\,H_2 +\kappa_{12}^{\prime}\, H_1^{\dag}\,H_2 H_2^{\dag}\,H_1+\kappa_{13}\, H_1^{\dag}\,H_1 H_3^{*} H_3 + \kappa_{23}\, H_2^{\dag}\,H_2 H_3^{*} H_3 \nonumber \\
&-&\frac{\mup}{\sqrt{2}}\, \left( H_1^{\dag}\,H_2 H_3 + H_2^{\dag}\,H_1 H_3^{*}\right)\,,\label{VSB} \\
\mathcal{V}_{\rm DM}&=& \mu_{S}^2\, S^{*} S + \lam_{S}\, (S^{*} S)^2+\Fs_{1}\, H_1^{\dag}\,H_1 S^{*} S +\Fs_{2}\, H_2^{\dag}\,H_2 S^{*} S + \Fs_{3}\, H_3^{*} H_3 S^{*} S \nonumber \\
&+&  h\, S^2 H_1^{\dag}\,H_2 + h^{*}\, S^{* 2} H_2^{\dag}\,H_1 
 - \frac{\mupp}{\sqrt{2}} (S^2 H_3^{*} + S^{* 2} H_3)\,.
 \label{VDM}
\eea
Through rotations of the scalar fields, all parameters but $h$ can be made real, while the dimensional parameters are assumed positive. The parameter $h$ is in general complex, but we will assume in the following that $h$ is real. 

The two scalar doublets $H_{1,2}$ and the complex scalar singlet $H_{3}$ are responsible for the breaking of $SU(2)_{W}\times U(1)_{Y}\times \left[U(1)_{B-\tilde{L}}\right]$ down to $U(1)_{em}\times \left[\mathcal{Z}_{2}\right]$.
Given the charges of $H_{2}$ and $H_{3}$, the discrete $\mathcal{Z}_2$ 
emerges as a remnant symmetry of the global $U(1)_{B-\tilde{L}}$ after EWSB.
Among the ten real scalar degrees of freedom, three of them are eaten through the Higgs mechanism, leaving a spectrum of seven physical scalars: two charged particles, $H^{\pm}$, two $CP$ odd neutral scalars, $A^{0}$ and the massless Majoron $\mathcal{J}$ \cite{Majoron}, and three $CP$ even neutral scalars, $h^{0},H^{0}$ and $h_{A}$. We derive in the following subsections 
some constraints  on the scalar sector parameter-space. An exhaustive phenomenological study, although of great interest, is beyond the scope of this work. 

The minimization of the scalar potential with respect to $H_{1}$, $H_{2}$ and $H_{3}$ vevs enforces three tree-level relations, that we use to define the quadratic terms $\mu_{i}$. Indeed, by parametrizing the Brout-Englert-Higgs fields and $S$ as
\bea
H_{k}&=&\left(H_{k}^{+}, \frac{v_{k}+h_{k}+\ii \,a_{k}}{\sqrt{2}}\right)^{T}\,,\quad {k=1,2}\,,\\
H_{3}&=&\frac{v_{3}+h_{3}+\ii \,a_{3}}{\sqrt{2}}\,,\quad S=\frac{S_{0}+\ii \,S_{1}}{\sqrt{2}}\,,
\eea
with
\bea
\left\langle H_{i} \right\rangle = \frac{v_{i}}{\sqrt{2}}\, \quad {\rm and }\quad \left\langle S \right\rangle = 0\,,
\eea
we get the extremum conditions 
\bea
\frac{\partial \mathcal{V}_{\rm SB}}{\partial v_{i}}=0\leftrightarrow \mu_{i}^{2}=\frac{1}{2}\left(v_{j}^{2} \tilde{\kappa}_{ij}+v_{k}^{2} \kappa_{ik}\right)+2\,v_{i}^{2}\lam_{i}-\frac{v_{j}v_{k}\mu^{\prime}}{2 v_{i}}\,,\quad i,j,k=1,2,3\,,\label{mincond}
\eea
where $\tilde{\kappa}_{12}=\kappa_{12}+\kappa^{\prime}_{12}$ and $\tilde{\kappa}_{ij}=\kappa_{ij}$ elsewhere.
The extremum obtained in (\ref{mincond}) is an absolute minimum provided the Hessian of $\mathcal{V}_{\rm SB}$ is positive definite.
Boundedness from below of the scalar potential requires the quartic couplings $\lam_{k}$ to be positive, as well as a non-trivial relation among the couplings. Notice that, 
since both $H_1$ and $H_2$ are charged under $SU(2)_{W}\times U(1)_{Y}$, they both contribute to the masses of the SM gauge bosons. 

Among the numerous parameters of $\mathcal{V}_{\rm SC}$, it is worth to emphasize the role of the trilinear 
coupling $\mup$. In \cite{Ma:2000cc,Grimus:2009mm}, a two-Higgs doublet model was built invariant under a $U(1)$ global symmetry, explicitly broken by a term $\propto \mu^{2}\phi_{1}^{\dagger}\phi_{2}$. Such term, for $\mu \ll v$ induces a type-II seesaw among the scalar vevs of $\phi_{1}$ and $\phi_{2}$:
$\langle \phi_{i}\rangle\ll \langle \phi_{j}\rangle$, $i\neq j$.
 As noted in \cite{Grimus:2009mm}, such  explicit breaking can be circumvented by the introduction of an additional scalar, say $\phi_{3}$, whose vev generates the required
 term: $\mu^{2} = \mu^{\prime} \langle \phi_{3}\rangle$. It is exactly along those lines that we build our scalar potential. Indeed, provided that 
 $\mup$ in (\ref{VSB}) is suppressed, $\mup\ll 1$ GeV, the minimization of $\mathcal{V}_{\rm SC}$ admits two possible hierarchical patterns for the vevs: $v_{3}\ll v_{2,1}$ and
 $v_{2}\ll v_{3,1}$.  As we will show below, only the latter is physically viable.
 One may wonder about the naturalness of such a suppressed mass parameter $\mup$. Let us stress that 
 very small values of $\mup$ are actually technically natural. Indeed, this term, as well as the couplings $h$ and $y_{2}$, are all terms linear in $H_2$. 
 By setting them to zero, one actually enlarges the symmetry group by an extra $U(1)$ factor. Therefore, small values of these parameters are natural, 
 in the 't Hooft sense \cite{'tHooft:1979bh}.

\mathversion{bold}
\subsection{$CP$ odd neutral scalars: $A^{0}$ and $\mathcal{J}$}\label{CPodd}
\mathversion{normal}

Three $CP$ odd neutral scalar fields arise from the spontaneous breaking of the  
electroweak symmetry: one pseudo-scalar $Z_{L}$, the longitudinal polarization
of the gauge boson $Z$, one massive pseudo-scalar $A^0$ 
and the massless Goldstone mode, associated with the spontaneous breaking
of the global symmetry $U(1)_{B-\tilde{L}}$, the Majoron 
$\mathcal{J}$.~\footnote{The Majoron $\mathcal{J}$ is exactly massless in our setup. 
Notice that $\mathcal{J}$, being the Goldstone boson of a spontaneously broken global symmetry, 
may acquire a mass through gravitational effects, as shown in \cite{Giddings:1988cx,Akhmedov:1992hi}. 
However, we will not consider this possibility in the following.}

The mass eigenstates are obtained by the basis transformation
\bea
\left(\ba{c} a_{1}\\a_{2}\\a_{3}\ea\right)=R_{PS}\,\left(\ba{c} Z_{L}\\\mathcal{J}\\A^{0}\ea\right)\,,
\eea
where $R_{PS}$ is a $3\times 3$ orthogonal matrix
\bea
R_{PS}\;=\;
\left(
\begin{array}{ccc}
 \cos(\beta) &  \sin^{2}(\beta)\,\Delta & -\sin(\beta)\,\tan(\gamma)\,\Delta \\
 \sin(\beta) & -\sin(\beta)\,\cos(\beta)\,\Delta & \cos(\beta)\,\tan(\gamma)\,\Delta \\
 0 & \tan(\gamma)\,\Delta & \sin(\beta)\,\Delta
\end{array}
\right)\,,
\eea
with $\Delta = \cos(\gamma)\,/ \sqrt{1-\cos(\gamma)^2 \cos(\beta)^2}$ and the mixing angles $\beta$ and $\gamma$ are by definition
\bea\label{angdef}
	\tan(\beta)\;=\;\frac{v_{2}}{v_{1}}\,,\quad\quad\quad\, \tan(\gamma)\;=\; \frac{v_{3}}{v_{1}}\,.
	\eea
The angles $\beta$ and $\gamma$ control the coupling of $\mathcal{J}$ to  SM fermions.
Indeed,  the interaction term relevant for the Majoron phenomenology is 
\bea
 -\mathcal{L}\supset \ii\,g_{\mathcal{J}ff}\;\,\ovl{f}\,\gamma_{5}\,f\, \mathcal{J}\,,
\eea
with
\bea\label{gjff}
g_{\mathcal{J}ff}\;\equiv\;\frac{m_{f}}{v }\,\sin(\beta)\,\tan(\beta)\,\Delta 
\eea
and $m_{f}$ is the fermion mass.
Strong constraints apply on these couplings, stemming from star cooling processes~\cite{AstroMaj}. 
In particular, the experimental upper limit on the cooling rate of white dwarfs 
implies: $|g_{\mathcal{J}ee}|\lesssim 10^{-12}$. 
Then, from (\ref{angdef}) and (\ref{gjff})  we obtain in the limit $\beta \ll 1$
\bea
	\beta \lesssim 7\times10^{-4}\sqrt{\tan(\gamma)} \,,
	\eea
implying $\, v_{2} \lesssim 0.2~{\rm GeV}\;\sqrt{v_{3} /v }$. 
As already stated at the beginning of this section, a hierarchical pattern of the type $v_{2}\ll v_{3} < v_{1} $ 
can be easily fulfilled, as the scale of $v_{2}$ is directly related to the dimensional parameter 
$\mup$:
\begin{eqnarray}
	v_{2}&\approx&\frac{v_{1}\, v_{3}\,\mup}{v_{1}^{2}\,\tilde{\kappa}_{12}\,+\,v_{3}^{2}\,\kappa_{23}\,-\,2\,\mu_{2}^{2}}\,,
\end{eqnarray}
which is suppressed compared to $v_{1,3}$ as $\mup$ can be naturally set to a scale much smaller than the EWSB scale.
The second physical pseudo-scalar, $A^{0}$, has mass
\bea
M_{A^{0}}^{2}\;=\;\mu^{\prime}\,\frac{v_{2}^{2}(v_{1}^{2}+v_{3}^{2})+v_{1}^{2} v_{3}^{2}}{2 \,v_{1} v_{2} v_{3}} \, \sim\,\mup\, \frac{v_{1}\,v_{3}}{2\,v_{2}}\,=\mup\,v\,\cos(\beta)\frac{\tan(\gamma)}{2\,\tan(\beta)}.
\label{massA0}
\eea
The actual value of $M_{A^0}$ depends on the ratio $\mup/ v_{2}$. Since the couplings of $A^{0}$ to the SM fermions  are $\sin(\beta)$ suppressed,
 $M_{A^0}$ and thus $\mup /v_{2}$ are unconstrained. 

\mathversion{bold}
\subsection{Charged scalars: $H^{\pm}$}
\mathversion{normal}

As in any two-Higgs doublet model, the charged scalar spectrum is composed of one physical field $H^{\pm}$ 
and the eaten longitudinal degree of freedom $W^{\pm}_{L}$. They are related to the interaction fields $H_{1,2}$ through
the orthogonal transformation:
\bea
\left(\ba{c} H_{1}^{+}\\H_{2}^{+}\ea\right)=\left(\ba{cc} \cos{(\beta)}&-\sin{(\beta)} \\ \sin{(\beta)}&\cos{(\beta)}\ea\right)\,\left(\ba{c} W_{L}^{+}\\H^{+}\ea\right)\,.
\eea
The charged scalar mass is given by
\bea
M_{H^{\pm}}^{2}=\frac{v^{2}}{2}\left( \frac{\tan(\gamma)}{\sin(\beta)}\frac{\mup}{v}-\K_{12}^{\prime} \right)\simeq \frac{M_{A^0}^{2}}{\cos(\beta)^{2}}-\frac{v^{2}}{2}\K_{12}^{\prime}\,,
\eea
where in the last expression we used the approximation given in (\ref{massA0}).
Since $M_{H^{\pm}}^{2}>0$, one requires $\kappa_{12}^{\prime}\lesssim 2 \,M_{A^{0}}^{2}/v^{2}$. 
An experimental lower bound on $M_{H^{\pm}}$ is obtained from $H^{\pm}$ pair production at LEP \cite{LEPch} and
the subsequent decays $H^{\pm}\to \tau^{\pm}\,\nu_{\tau}$ and $H^{\pm}\to c \,s$. 
For $m_{H^{\pm}}\leq m_{W}$, $H^{\pm}$ decays only to SM fermions, so the bound $M_{H^{\pm}}\gtrsim 78.6\GeV$ applies.

\mathversion{bold}
\subsection{$CP$ even neutral scalars: $h^{0}$, $H^{0}$ and $h_{A}$}
\mathversion{normal}

We introduce the $CP$ even mass eigenstates $h^{0}$, $H^{0}$ and $h_{A}$, which are
related to the interaction fields $h_{1,2,3}$ through the basis rotation: 
\bea
\left(\ba{c} h_{1}\\h_{2}\\h_{3}\ea\right)=R_{NS}\,\left(\ba{c} H^{0}\\h_{A}\\h^{0}\ea\right)\,,
\eea
In the limit $\mup,v_{2}\ll v_{3}< v_{1}$, the $CP$ even scalar mass matrix can be further simplified and $R_{NS}$ just consists in a rotation
of angle $\theta$ between the eigenstates $h^{0}$ and $H^{0}$. Moreover, at leading order in $\beta$, 
$h_{A}$ and the pseudo-scalar $A^{0}$ are degenerate in mass and both decouple from the other particles, so no constraints apply on $M_{h_{A}}$. 
Within this approximation and introducing as a shorthand 
\bea
m_{1}=v_{1}\,\sqrt{2\,\lam_{1}}\,,\quad m_{3}=v_{3}\,\sqrt{2\,\lam_{3}}\quad\text{and}\quad m_{13}=\sqrt{v_{1}\,v_{3}\,\K_{13}}\,, \nonumber
\eea
the masses of the neutral Higgs $H^0$ and $h^0$ and the mixing angle $\theta$ are given by the relations
\bea
M_{H/ h}^{2}=\frac{1}{2}\left(m_1^2 + m_3^2 \,\pm\, \sqrt{\left(m_{1}^2-m_{3}^2\right)^2\,+\,4\,m_{13}^4}\right)\,,\quad
\theta = {\rm Arctan}\left( \frac{m_{3}^2-m_{1}^2+\sqrt{\left(m_{3}^2-m_{1}^2\right)^2\,+\,4\,m_{13}^4}}{2\,m_{13}^2} \right)\,,
\eea
and $M_{H^{0}} \geq M_{h^{0}}$. The mixing angle $\vert \theta \vert$ takes values from zero  to $\pi/2$.
The couplings of $H^0$ ($h^0$) to SM particles are given by the SM Higgs ones times $\cos(\theta)$ ($\sin(\theta)$). For maximal $\theta\sim \pi/2$, $h^0$ couplings are unsuppressed compared to the SM case, so LEP-II bounds apply and $M_{H^0}\geq M_{h^0}\gtrsim 114.4 \GeV$~\cite{LEPII}. In the opposite case, with suppressed mixing angle $\vert \theta \vert \ll 1$, only $H^0$ get sizable couplings to the SM, and the former bound on $M_{H^0}$ still applies. Conversely, for $\vert \theta \vert  \ll 1$, LEP-II bounds are  rather weak in constraining the mass of the lightest Higgs. Notice that $h^0$ contributes to the invisible $Z$ decay; however, the $Z-h^{0}-J$ coupling is $\beta^{4}$ suppressed: this contribution is negligible and no relevant constraints apply on $M_{h^0}$ from this decay. Nevertheless, for $\sin^{2}(\theta) \gtrsim 0.1$, LEP-II bounds imply $M_{h^0} \gtrsim 80 \GeV$. In the following, we assume the conservative limit $M_{h^0} \gtrsim 114.4 \GeV$, which is valid for all values of $\theta$. 

\subsubsection*{An almost invisible Higgs boson}

As occurs in models with multiple scalars, the Higgs bosons may decay invisibly.
In our scenario, both $H^0$ and $h^{0}$ can decay into two Majorons, thus precluding their detection at present particle colliders,
LHC included. 

The total decay widths of $H^{0}$ and $h^0$ are given by
\bea\label{DecHiggs}
\Gamma(H/h)\simeq\frac{1}{8\pi}\sum_{ij} \frac{\kappa\left(M_{H/h},M_{i},M_{j}\right)}{2\,M_{H/h}^{3}}\vert \mathcal{M}_{i\,j}\vert^{2}\, \Theta\left(M_{H/h}^{2}-\left(\,M_{i}+M_{j}\right)^2\right)\,.
\eea
In the equation given above the kinematical factor is  $\kappa(x,y,z)\equiv\sqrt{(x^2-(y+z)^2)(x^2-(y-z)^2)}$. 
We consider below for simplicity only tree-level two-body decays into identical final states.
The decay probabilities of $H^0$ and $h^0$ to neutral scalars are proportional to the norms of the trilinear couplings, which at zeroth order in $\beta$ read:
\bea
\left\vert \mathcal{M}^{H/h}_{\mathcal{J}} \right\vert^{2}&=& v^{2}\,\left\vert \lam^{H/h}_{\mathcal{J}\,\mathcal{J}}\right \vert^{2},\, 
\lam^{H/h}_{\mathcal{J}\,\mathcal{J}} = \cos(\beta)\,\K_{13} \left\lbrace \ba{c} \cos(\theta)\\ \sin(\theta) \ea\right. -2\cos(\beta)\tan(\gamma)\,\lam_{3}\left\lbrace  \ba{c} \sin(\theta)\\ {\rm -}\cos(\theta) \ea\right. , \label{HHdecay1}\\
\left\vert \mathcal{M}^{H/h}_{H^{\pm}} \right\vert^{2}&=&v^{2}\,\left\vert \lam^{H/h}_{H^{+}H^{-}} \right\vert^{2},\, 
\lam^{H/h}_{H^{+}H^{-}} =
\cos(\be)\,\kappa_{12} \left\lbrace\ba{c} \cos(\theta) \\  \sin(\theta)\ea \right.-\cos(\beta)\tan(\gamma)\,\K_{23}\left\lbrace \ba{c}\sin(\theta) \\ {\rm -}\cos(\theta) \ea\right. , \label{HHdecay1b}\\
\left\vert \mathcal{M}^{H/h}_{h_{A}} \right\vert^{2}=\left\vert \mathcal{M}^{H/h}_{A}\right \vert^{2}&=& v^{2}\,\vert \lam^{H/h}_{A\,A} \vert^{2},\,
 \lam^{H/h}_{A\,A} =\cos(\be)\,\tilde{\K}_{12} \left\lbrace\ba{c} \cos(\theta) \\  \sin(\theta)\ea \right.-\cos(\beta)\tan(\gamma)\,\K_{23}\left\lbrace \ba{c}\sin(\theta) \\ {\rm -}\cos(\theta) \ea\right. , \label{HHdecay2}\\
\left \vert \mathcal{M}^{H/h}_{S} \right\vert^{2}&=& v^{2}\,\left\vert \lam^{H/h}_{S\,S} \right\vert^{2},\,
 \lam^{H/h}_{S\,S} =\cos(\be)\,\Fs_{1} \left\lbrace\ba{c} \cos(\theta) \\  \sin(\theta)\ea \right.-\left(\Fs_{3}\cos(\beta)\tan(\gamma)\,-\frac{\mupp}{v}\right)\left\lbrace \ba{c}\sin(\theta) \\ {\rm -}\cos(\theta) \ea\right. \label{HHdecay3}
\eea
The decay probabilities of $H^{0}$ and $h^{0}$ into SM particles are similar to the SM ones,  see $e.g.$ \cite{DjouadiSM}, modulo a dependence
on the mixing angles $\theta$ and $\beta$. At tree level, for the decay probability into fermions, we have
\bea
\left\vert \mathcal{M}^{H/h}_{f} \right\vert^{2} &=&2\,N_c\,\left(\frac{m_f}{v}\right)^{2}\,M_{H/h}^{2}\, \left(1-4\,\frac{m_{f}^{2}}{M_{H/h}^{2}}\right) \left\lbrace\ba{c} \cos^{2}(\theta) \\ \sin^{2}(\theta) \ea \right. \,,
\eea
where $N_c$ is the number of colors and $m_{f}$ is the fermion mass. 
The tree-level $H^{0}/h^{0} \to W^+\,W^-$ decay probabilities depend on
\bea
\left\vert \mathcal{M}^{H/h}_{W^\pm}\right \vert^{2} &=& \frac{g_{W}^{4}}{16}\,v^{2}\,\left\vert\lam^{H/h}_{W}(M_{H/h}) \right\vert^{2}\,,\\
\lam^{H/h}_{W}(m)&=&\cos(\beta)\,\frac{m^{2}}{M_W^2}\sqrt{1-4\frac{M_W^2}{m^{2}}+12\,\frac{M_W^4}{m^4}}\,
\left\lbrace \ba{c} \,\cos(\theta) \\ \,\sin(\theta) \ea\right. \,,\label{lww}
\eea
where $M_{W}$ is the $W$-boson mass and $g_{W}$ is the weak gauge coupling constant. A similar expression holds for $H^0/h^0$ decays into pairs of $Z$ bosons.
\begin{figure}[t!]
\begin{center}
\includegraphics[width=0.8\textwidth]{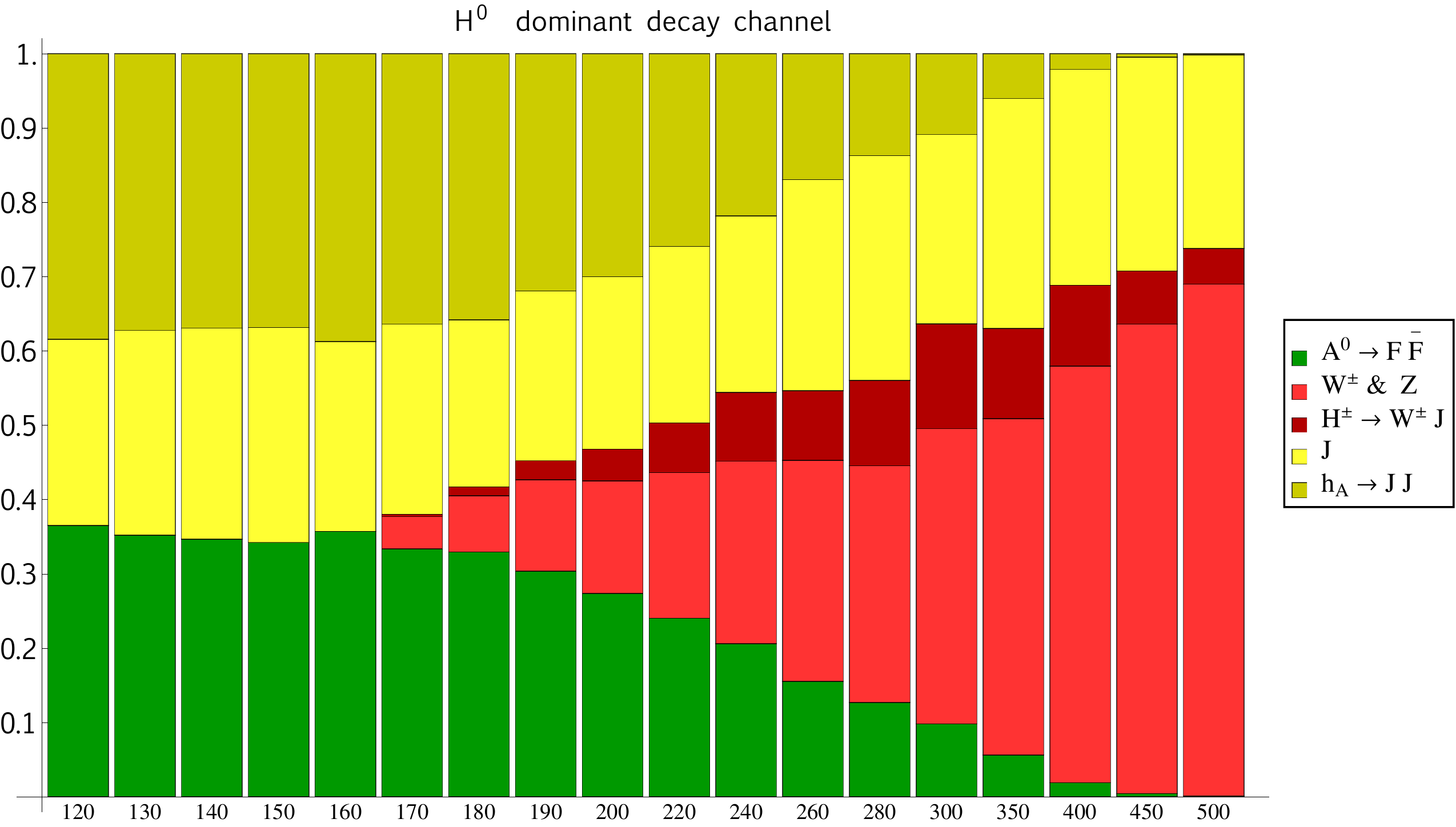}
\includegraphics[width=0.8\textwidth]{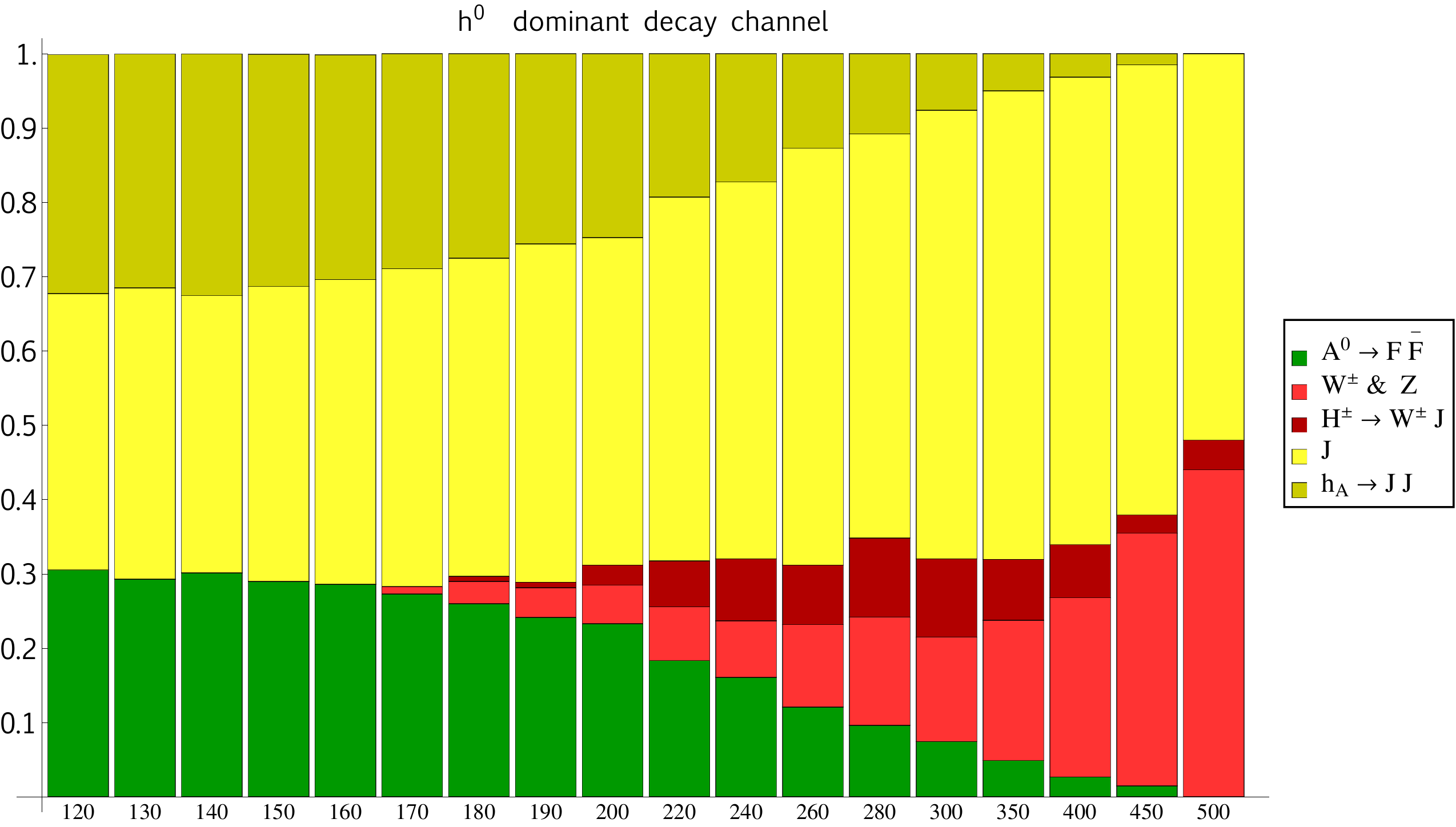}

\caption{Dominant $H^0$ (top) and $h^0$ (bottom) Higgs decay channels in function of their masses.}
\label{GrapheHiggsBRSM}
\end{center}
\end{figure}

From the expressions above, we can estimate the invisible branching ratio of the Higgs bosons. 
First of all, obviously, the smaller the quartic portal couplings, the smaller the invisible Higgs decay widths. Second, the decays into $h_{A}$ or $A^0$, even if equal at leading order in $\beta$, do not have a similar impact on Higgs searches at colliders. 
Indeed, the pseudo-scalar $A^0$ eventually decays almost exclusively into SM fermions:  $H^0/h^0 \to A^0\,A^0$ can thus be considered as a visible channel. Conversely, $h_{A}$ essentially decays into Majorons.

In the low mass region, $M_{H/h}\lesssim 2\,M_{W}$, neglecting the masses of the decay products, the Higgs invisible to visible decay width ratios are
\bea
\frac{\Gamma(H/h \to {\rm inv}) }{\Gamma(H/h \to {\rm SM})} \simeq \frac{\Gamma(H/h \to \mathcal{J}\mathcal{J})+\Gamma(H/h \to h_{A}\,h_{A})+\Gamma(H/h \to S\,S)}{\Gamma(H/h \to b\,\ovl{b})+\Gamma(H/h \to A^{0}\,A^{0})} \,, 
\eea
provided $M_{h_{A}}\simeq M_{A^0} \lesssim M_{H/h}/2$. The decay width to $b$-quarks is Yukawa suppressed, so the ratio above is simplified to
\bea
\frac{\Gamma(H/h \to {\rm inv}) }{\Gamma(H/h \to {\rm SM})} \simeq 1+\frac{\Gamma(H/h \to \mathcal{J}\mathcal{J})+\Gamma(H/h \to S\,S)}{\Gamma(H/h \to A^0\,A^0)}\,. 
\eea
Consequently, for low masses, both $H^0$ and $h^0$ mostly decay invisibly. 

 In the  high mass regime, $M_{H/h}Ê\gtrsim 2\,M_W$, under the approximation that the Majoron channel constitutes the main invisible decay and the visible channel is mostly due to decays to gauge bosons, we have
 \bea
\frac{\Gamma(H/h \to {\rm inv}) }{\Gamma(H/h \to {\rm SM})} \simeq \frac{\Gamma(H/h \to \mathcal{J}\mathcal{J})}{\Gamma(H/h \to W^+\,W^-)+\Gamma(H/h \to Z\,Z)}\propto 
\frac{16}{g_{W}^4}\frac{M_{W}^{4}}{M_{H/h}^4} \left( \K_{13} \,\mp 2\,\lam_{3}\,\tan(\gamma)\,\left\lbrace \ba{c} \,\tan(\theta) \\ \,1/\tan(\theta) \ea\right.\right)^2\,. 
\eea
From the previous estimate we  infer that, for a maximal $\theta\sim \pi/2$, the heaviest Higgs boson, $H^{0}$, decays prevalently into two Majorons, thus forbidding its detection  at current collider searches. The opposite occurs for the lightest $CP$ even scalar $h^{0}$. On the other hand, 
for higher values of $M_{H^{0}}$ ($M_{h^{0}}$) and a sufficiently small (large) mixing angle $\theta$, the visible decay rate of $H^{0}$ ($h^{0}$) becomes sizable. It dominates for very heavy Higgs bosons.

In Fig.~\ref{GrapheHiggsBRSM} we display the frequency at which the $H^0$ and $h^0$ decays channels are the dominant ones, displayed in the top and bottom panels, respectively. 
In order to produce this plot, we use the Higgs decay branching fractions computed by the program micrOMEGAs \cite{micromegas}, 
that we also use to study the Dark Matter sector, as discussed in Section~\ref{DMSec}.~As expected, we see 
from Fig.~\ref{GrapheHiggsBRSM} that above the $W$ threshold, the heavier the Higgs bosons  the larger their visible decay rates.~\footnote{Notice that
 we only consider the two-body decay widths $H^{0}/h^0\to W^{+}\,W^{-}$. However, in the SM the tree-body  decays through off-shell $W$ actually dominate
for $M_{H/h} \gtrsim 135 \GeV$, cf. \cite{DjouadiSM}, in which case the Higgs visible decay channels should prevail here as well.}~Conversely, 
in the low mass regime the Higgs bosons are clearly unobservable as we explained above.

\section{Dark Matter}\label{DMSec}

We discuss in this section the third building-block of our model: the existence of a viable Dark Matter candidate. Below the EWSB scale, the complex scalar $S$ is split 
 into two real components $S_0$ and $S_1$, the lightest one being the DM. Real scalar singlets provide the simplest DM candidates, for which 
 a large literature exists \cite{SingletDM}. In our model, we shall stress two important aspects: first, the stability of DM is not an \textit{ad-hoc} prescription, but results from the remnant $\mathcal{Z}_2$, $S_{0}$ or $S_{1}$ being the lightest particle odd under this discrete symmetry; second, we emphasize again that introducing the scalar $S$ not only provides a DM candidate, but 
 is also necessary in our leptogenesis scenario. 

The masses of the two real components of $S$ are:
\bea
m_{S_{0(1)}}^{2}&=&\mu_{S}^{2}+\frac{1}{2}\left(\Fs_1\,v_{1}^{2}+\Fs_2\,v_{2}^{2}+\Fs_3\,v_{3}^{2}\right)
\,\pm \,\left(h\,v_{1}\,v_{2}\,-\,\mupp\, v_{3}\right)\,.
\label{DMmass}
\eea
The mass splitting in this case is controlled by the parameters $h$ and $\mu^{\prime\prime}$. 
However, since $v_{2} \ll v_{3}$, the latter term dominates and $m_{S^0} \leq m_{S_{1}}$ for positive $\mupp$.~\footnote{In the following we will however denote by $S$
the DM candidate. The heavier state will decay to DM plus Majoron.}

As seen from $\mathcal{V}_{\rm DM}$, eq.~(\ref{VDM}), $S$ has several portal  couplings to the Higgs fields, 
implying many annihilation channels~\cite{Portal}. Like in most of the singlet scalar DM scenarios, $S$
easily gets a thermal relic abundance in agreement with cosmological requirements.

\subsection{Relic density}

The DM annihilation cross-section can generically be written as
\bea
\sigma\,v\sim \frac{\lam_{eff}^{2}}{m_{S^2}}\,,
\eea
where the effective coupling $\lam_{eff}$ indicates that each annihilation channel receives in general several contributions. When $S$ annihilates into scalars, the cross-section is the (coherent) sum of the contact term interaction, for which $\lam_{eff} \propto \Fs_{i}$, cf. eq.~(\ref{VDM}), and of scalar-mediated interactions, where $\lam_{eff}$ depends on the different trilinear scalar couplings, such as $\lam^{H/h}_{S\,S}$ introduced in the previous section. 

For light DM, that is $m_{S}\lesssim M_{W} $, $S$ mostly annihilates to Majorons, as well as to pairs of $h_A$ or $A^0$, granted the latter are light enough. Notice that the annihilation cross-sections into pairs of $h_A$ and $A^{0}$ coincide at zeroth order in $\beta$. 
\begin{figure}[htb!]
\begin{center}
\includegraphics[width=0.7\textwidth]{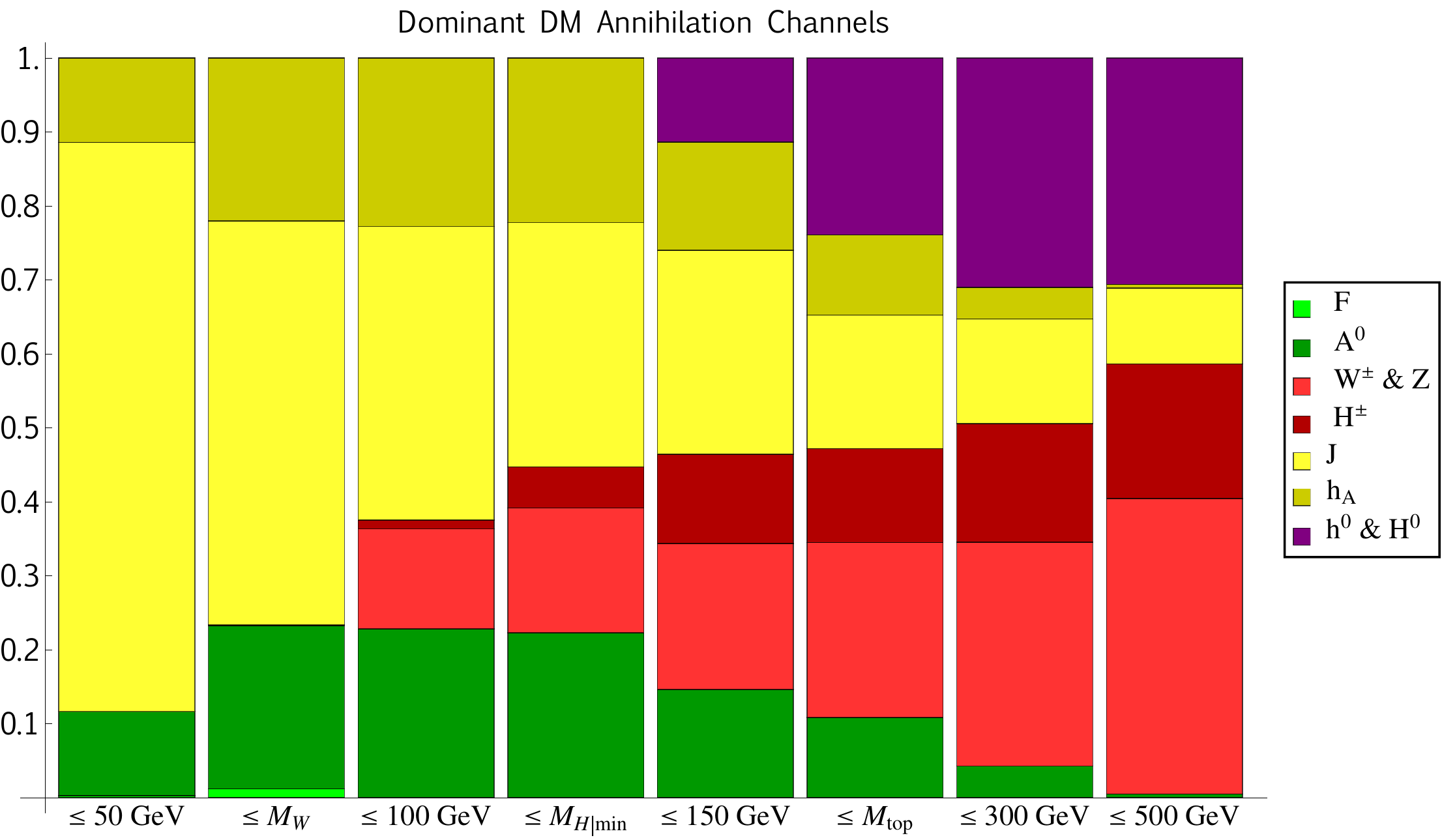}
\caption{Main Dark Matter annihilation channels for different Dark Matter mass ranges.}
\label{GrapheDMann}
\end{center}
\end{figure}\\
For heavier DM, new annihilation channels are open. In the case $m_{S}>M_{W}$,  $S$ can annihilate into pairs of $W^\pm$ through $h^0$ and $H^0$ $s$-channel (the $h_{A}$ mediation is $\beta^2$ suppressed):~\footnote{For simplicity, the widths of the Higgs fields have been neglected in eq.~(\ref{AnnWW}), although they are taken into account in the numerical evaluation.}
\bea\label{AnnWW}
\sigma v=\frac{g_{W}^4}{16\,\pi\,m_S^2} \left(\frac{\,v^{2}\,\lambda_{S\,S}^{H}\,\lam^{H}_{W}(2\,m_S)}{M_{H^0}^2-4\,m_S^2}+\frac{\,v^{2}\,\lambda_{S\,S}^{h}\,\lam^{h}_{W}(2\,m_S)}{M_{h^0}^2-4\,m_S^2}\right)^{2},
\eea
where $\lam^{H/h}_{W}(m)$ were introduced in eq.~(\ref{lww}). A similar expression holds for the annihilation into pairs of $Z$ bosons.
In the high mass range, $S$ may also annihilate into pairs of charged $H^\pm$, or to pairs of $CP$ even scalars $h^0$ and $H^0$.

Increasing the DM mass, the quartic couplings $\Fs_{i}$ which control the DM mass, eq.~(\ref{DMmass}), and the effective couplings $\lambda_{S\,S}^{H/h}$,
eq.~(\ref{HHdecay3}), should increase as well, so that the annihilation cross-section remains large enough, in order to obtain the observed DM relic abundance.

\subsection*{Numerical evaluation}

In order to accurately determine the relic abundance of $S$, we implemented our model in micrOMEGAs \cite{micromegas}, through the program FeynRules \cite{FeynRules}. We then performed a scan over the full scalar parameter-space, by assigning random values to the different couplings. All $\lam$ and $\K$ quartic couplings were varied from $10^{-4}$ 
up to the perturbative bound $4\pi$. The trilinear coupling $h$ was chosen between $10^{-6}$ and $10^{-2}$. The scalar masses were randomly varied from their experimental lower bounds, discussed in the previous section, up to 500 GeV. In particular, as regards the $CP$ even scalar masses, recall that we impose
the conservative bound $M_{H^{0}/h^{0}}\gtrsim 115$ GeV. We vary the mixing angle $\theta$ in the range: $0\leq \vert \theta \vert \leq\pi/2$.
For the unconstrained scalars $h_{A}$ and $A^0$, their (almost degenerate) mass was varied between $1\GeV$ and $100\GeV$.
 
The trilinear mass term $\mupp$ was scanned over in the range $(1$-$10^{2}) \GeV$, 
while $\mup$ typically took values between $10$ eV and $10$ MeV. Finally, $\mu_S$ was varied from $1 \GeV$ to $500\GeV$.

We demand the relic density of $S$ to account for all the DM abundance and to lie within the 
$3\sigma$ range of WMAP \cite{WMAP}:  
\bea
\Omega_{\rm DM}=0.229 \pm 0.045 \,.\nonumber
\eea
We illustrate the relative contributions of the different annihilation channels in Fig.~\ref{GrapheDMann}. 
Binning the DM mass range into intervals of interest, we present the frequency at which a given channel is the dominant one.
For example, before the $W$ channel is kinematically open, $i.e.$ for $m_S \leq M_W$, we see from Fig.~\ref{GrapheDMann} that $S$ annihilates only into
pairs of $\mathcal{J}$, $A^{0}$ and $h_{A}$. For heavier DM mass, new annihilation processes are possible.
In particular, annihilation into gauge bosons, charged scalars or $CP$ even scalars tend to be the dominant processes.
Notice that Fig.~\ref{GrapheDMann} only displays the frequency a given annihilation channel dominates in a given mass interval and not the relative contributions of the different channels.

\subsection{Direct detection constraints}

The Dark Matter can scatter on nucleons through scalar-mediated $t$-channels: 
the spin-independent (SI) elastic  cross-section receives contributions from both $h^{0}$ and $H^{0}$ exchange, according to:
\bea
\sigma_{n}^{SI}=\frac{1}{4\pi}\frac{\mu_{S,n}^{2}}{m_{S_0}^{2}}\,m_{n}^{2}\,f_{n}^{2}\left( \frac{\lam_{H^0}}{M_{H^{0}}^{2}}+\frac{\lam_{h^0}}{M_{h^{0}}^{2}}\right)^{2}\,.
\eea
In this expression, $\mu_{S,n}$ is the $S$-nucleon reduced-mass and $m_{n}$ the nucleon mass. The factor $f_n$ is the effective Higgs-nucleon interaction and  varies from  $0.14$ to $0.66$ \cite{FN}. The couplings $\lam_{H^{0}}$ and $\lam_{h^{0}}$ are given by
\bea\label{lamSI}
\lam_{H/h}=\frac{1}{\cos(\beta)}\,\lambda_{SS}^{H/h}\,\left\lbrace \ba{c} \cos(\theta) \\ \sin(\theta)\ea \right.
\eea
Assuming the conservative bound $M_{H^{0}/h^{0}}\gtrsim 115$ GeV, we see from the previous expression that for 
$\theta\approx 0~(\pi/2)$ the main contribution comes from $H^0$ ($h^{0}$) exchange
and $\sigma_{n}^{SI}$ is then mostly affected by  $\Fs_1$. Notice that, 
contrary to \cite{FXdm}, where the mixing suppression $\theta \ll 1$ was balanced by a very light scalar $h^{0}$ ($M_{h^{0}}\lesssim 1\GeV$), in the present scenario, taking $M_{h^{0}}$ above the LEP-II bound drastically forbids such an enhancement. 
\begin{figure}[t!]
\begin{center}
\includegraphics[width=0.7\textwidth]{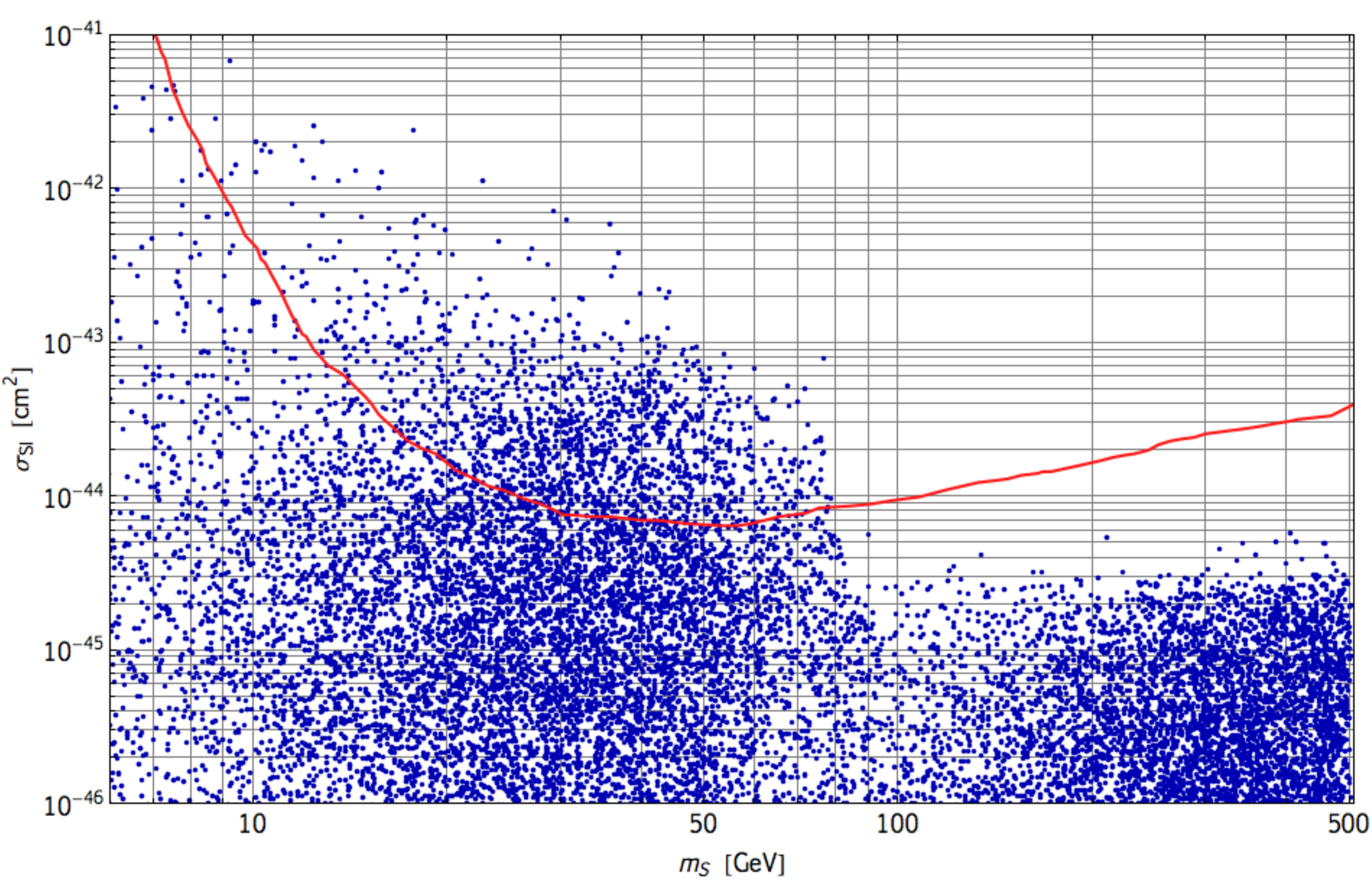}
\caption{Spin-independent cross-section against $m_S$: the blue points are the model predictions which provide the required relic density. The red line represents XENON100 results, extracted from \cite{XENON100}.}
\label{GrapheDMDD}
\end{center}
\end{figure}
In the limit of small mixing angle $\theta$, assuming $f_{n}\sim 1/3$, the SI  elastic cross-section can be roughly expressed as
\bea
\sigma_{n}^{SI}\simeq \frac{1}{\pi}\frac{m_{n}^{4}}{m_{S}^{2}}\,f_{n}^{2} \left(\frac{ \Fs_1 }{M_{H^{0}}^{2}}\right)^{2}\,\sim 6\times 10^{-44}\,{\rm cm}^{2}\left(\frac{m_{S}}{100\GeV}\right)^{-2}\,\left(\frac{M_{H^0}}{120\GeV}\right)^{-4}\left(\frac{\Fs_1}{0.1}\right)^{2}\,,
\eea
which shows that $S$ can easily saturate current direct-detection bound for electroweak scale DM~\cite{DMdd1}-\cite{XENON100}. 
As we saw in the previous subsection, since the annihilation cross-section scales as $(\Fs_{k} / m_{S})^{2}$, the couplings $\Fs_{k}$ should be sizable for large DM masses, otherwise $S$ relic density would overclose the Universe. This, in turn, implies that for heavy DM the scattering cross-section on nucleons is almost independent of the DM mass, cf. eq.~(\ref{lamSI}).

The dependence of $\sigma_{n}^{SI}$ on $m_{S}$ in the low and high DM mass regimes is manifest
in Fig.~\ref{GrapheDMDD}. In this plot
we compare the model predictions (blue dots)  for $\sigma_{n}^{SI}$ with XENON100 results \cite{XENON100} (red curve). 
We can see that while only a small region of the parameter-space is already excluded by current data, the next generation of direct-detection experiments would probe a large part of it~\cite{DMprospects}. Notice in particular that in the low mass regime high cross-sections can be reached, due to non-suppressed $\Fs_{1,3}$ couplings. A light DM with large $\Fs_{1,3}$ couplings is possible through a partial cancellation in eq.~(\ref{DMmass}), which depends on the value of the parameter $\mupp$.

\section{Conclusions}

In this paper we study  a seesaw extension of the Standard Model 
based on a global $U(1)_{B-\tilde{L}}$ symmetry group, where $\tilde{L}$ can be thought as
a generalized lepton charge. This global symmetry is  spontaneously broken
at the electroweak scale. Suitable scalar and fermion representations
are added to the SM particle content so that 
a tiny Majorana mass for active neutrinos is naturally generated,
in agreement with neutrino oscillation experiments. 
More specifically,  an extra Higgs doublet $H_{2}$ and a Higgs singlet $H_{3}$ are added to the SM, together with a heavy Dirac fermion $N_{D}$.
The lepton doublets and $N_{D}$ interact through neutrino Yukawa couplings which can violate the lepton number. When  $N_{D}$ mass is set at the TeV scale, the model realizes a UV-completion of the inverse-seesaw mechanism.

We show that, with the addition of two extra SM-singlets in the model, 
a Majorana fermion $N_{3}$ and a complex scalar $S$,
it is possible to explain quantitatively  both the observed baryon asymmetry of the Universe 
through an original leptogenesis mechanism and the Dark Matter relic abundance.

Leptogenesis in this model is implemented in  two steps: 
first an asymmetry in $N_{D}$ and $S$ is generated by the 
out-of-equilibrium decays of  $N_{3}$.
In a second step this asymmetry is converted in a non-zero lepton charge due to fast neutrino Yukawa interactions.
The latter constitute a link between leptogenesis and neutrino mass generation.
We solve numerically the Boltzmann equations relevant for this two-step leptogenesis scenario and show that
the observed amount of baryon asymmetry is easily achieved.
We concentrate the discussion  on a TeV scale scenario, and show that, provided neutrino mass constraints are fulfilled, no lower-bound on the mass of $N_{D}$ is imposed by the requirement of  a successful leptogenesis. However, this scenario of leptogenesis is also viable at much larger scales.
An important feature of this mechanism is that the source and damping terms do not depend on the same couplings, therefore  large $CP$ asymmetries can be obtained even in the regime of weak washouts.

In the second part of the paper, we analyze in detail the mass spectrum
of the model and provide constraints on the parameter-space arising from low-energy physics. In particular, we show that 
the presence of a massless Majoron, which corresponds to the 
Goldstone boson associated with the spontaneous breaking of the global $U(1)_{B-\tilde{L}}$ symmetry, has an important impact on Higgs boson searches.
Indeed, light Higgs scalars $H^{0}$ and $h^0$,  $M_{H^{0}/h^0}\lesssim 140$ GeV, would mainly decay into  pairs of Majorons, thus making difficult their observation at colliders, LHC included.

Concerning the scalar field $S$, after the breaking of the electroweak symmetry, the lightest component of $S$ remains stable, due to the presence of a remnant $\mathcal{Z}_{2}$ symmetry,
and provides a viable candidate for Dark Matter. Its mass is unconstrained and can take values as light as few GeV up to few TeV. Numerous annihilation channels are present, allowing the relic DM density to be consistent with cosmological observations. We study the possible signatures of DM in direct detection experiments, and show that while the current constraints exclude already a part of the parameter-space, a large region may be probed by the next generation of detectors.

This model explains in a common framework three main experimental issues: neutrino mass generation, the baryon asymmetry of the Universe and the Dark Matter relic density. 
Many observables are predicted, but their measurements probe uncorrelated sectors, making this minimal extension difficult  to falsify.

\section*{Acknowledgements}
The authors acknowledge N.~Bernal for early stage discussions and S.~Palomares-Ruiz for useful comments.
The work of F.X.J.M. and E.M. is supported by
Funda\c{c}\~{a}o para a Ci\^{e}ncia e a
Tecnologia (FCT, Portugal) through the projects
PTDC/FIS/098188/2008,  CERN/FP/116328/2010
and CFTP-FCT Unit 777,
which are partially funded through POCTI (FEDER).

\appendix

\section{Interaction Lagrangian of the Model}\label{AppLag}

The full interaction Lagrangian of the model is:

\begin{eqnarray}\label{Linter}
	\mathcal{L}_{\rm int}&=&\mathcal{L}_{\rm int}^{\rm SM}\,-\,\mathcal{V}_{\rm SB}\,-\,\mathcal{V}_{\rm DM}\,-\,M\,\overline{N}_{D}N_{D}\,
	-\,\frac{1}{2}M_{3}\,\overline{N}_{3}\,N_{3}^c	
	\,\\
	&&-\left( y_{1}^{i}\,\overline{N}_{D}\,\widetilde{H}_{1}^{\dagger}\,\ell_{i}\,
	+\,y_{2}^{j}\,\overline{N}_{D}^{c}\,\widetilde{H}_{2}^{\dagger}\,\ell_{j}
	\,+\,\frac{\alpha}{\sqrt{2}}\,H_{3}\,\overline{N}_{D}\,N_{D}^{c}\,+g\,S\,\overline{N}_{D}\,N_{3}\,+\,{\rm h.c.}\right)\,, \nonumber	
\end{eqnarray}
where $\mathcal{L}_{\rm int}^{\rm SM}$ is the Yukawa interaction Lagrangian 
of the Standard Model and
\bea
\mathcal{V}_{\rm SB}&=&-\mu_{1}^2\, H_1^{\dag}\,H_1 + \lam_{1}\, (H_1^{\dag}\,H_1)^2 - \mu_{2}^2\, H_2^{\dag}\,H_2 + \lam_{2}\, (H_2^{\dag}\,H_2)^2 - \mu_{3}^2\, H_3^{*} H_3 + \lam_{3}\, (H_3^{*} H_3)^2\nonumber \\  &+& \kappa_{12}\, H_1^{\dag}\,H_1 H_2^{\dag}\,H_2 +\kappa_{12}^{\prime}\, H_1^{\dag}\,H_2 H_2^{\dag}\,H_1+\kappa_{13}\, H_1^{\dag}\,H_1 H_3^{*} H_3 + \kappa_{23}\, H_2^{\dag}\,H_2 H_3^{*} H_3 \nonumber \\
&-&\frac{\mup}{\sqrt{2}}\, \left( H_1^{\dag}\,H_2 H_3 + H_2^{\dag}\,H_1 H_3^{*}\right)\,,\\
\mathcal{V}_{\rm DM}&=& \mu_{S}^2\, S^{*} S + \lam_{S}\, (S^{*} S)^2+\Fs_{1}\, H_1^{\dag}\,H_1 S^{*} S +\Fs_{2}\, H_2^{\dag}\,H_2 S^{*} S + \Fs_{3}\, H_3^{*} H_3 S^{*} S \nonumber \\
&+&  h\, S^2 H_1^{\dag}\,H_2 + h^{*}\, S^{* 2} H_2^{\dag}\,H_1 
 - \frac{\mupp}{\sqrt{2}} (S^2 H_3^{*} + S^{* 2} H_3)\,.
\eea

\section{Computation of the $CP$ Asymmetry}\label{CompCPAsymm}

\noindent The relevant interaction Lagrangian which is involved in the generation of the $CP$ asymmetry in the 
out-of-equilibrium decays of the Majorana neutrino $N_{3}$ is the following:
\begin{eqnarray}
	-\mathcal{L}_{int} &\supset&
				\frac{\alpha}{\sqrt{2}}\,H_{3}\,\overline{N}_{D}\,N_{D}^{c}
				\,+\,g\,S\,\overline{N}_{D}\,N_{3}\,-\,\frac{\mu^{\prime\prime}}{\sqrt{2}}S^{2}H_{3}^{*}\,+\,{\rm h.c.}\,,\nonumber
\end{eqnarray}
where $N_{D}^{c}\equiv C\overline{N}_{D}^{T}$, $N_{3}\equiv N_{3}^{c}\equiv C\overline{N}_{3}^{T}$. The $CP$ asymmetry in the decays of $N_{3}$ is defined as:
\begin{eqnarray}
	\epsilon_{CP} &\equiv & -\frac{\Gamma\left(N_{3}\to\overline{N}_{D}+S\right)-\Gamma\left(N_{3}\to N_{D}+\overline{S}\right)}
						{\Gamma\left(N_{3}\to\overline{N}_{D}+S\right)+\Gamma\left(N_{3}\to N_{D}+\overline{S}\right)}\nonumber\\\label{CPAsymm}\\
	 &=&- \frac{\text{Im}\left\{\int\, d\tilde{\Pi}_{N,S}\, 
	\mathcal{M}^{(0)}(N_{3}\to\overline{N}_{D}\,S)^{*}\,\sum\limits_{\{n\}}\,\int\,d\tilde{\Pi}_{\{n\}}\,
	\mathcal{M}^{(0)}(N_{3}\to\{n\})\,\mathcal{M}^{(0)}(\{n\}\to\overline{N}_{D}\,S)\right\}}
	{\int\,d\tilde{\Pi}_{N,S}\,\left|\mathcal{M}^{(0)}(N_{3}\to\overline{N}_{D}\,S)\right|^{2}}\,,\nonumber
\end{eqnarray}  
where $\sum_{\{n\}}$ indicates the sum over all possible on-shell states in the loop of Fig.~\ref{CPAsymmGraph}, while the phase 
space factor in the integral is, in general          
\begin{equation}
	d\tilde{\Pi}_{n_{1},\ldots,n_{k}}\;\equiv\;\frac{d^{3}p_{n_{1}}}{(2\pi)^{3}2E_{n_{1}}}\cdot\ldots\cdot\
	\frac{d^{3}p_{n_{k}}}{(2\pi)^{3}2E_{n_{k}}}\,(2\pi)^{4}\,\delta^{(4)}\left(p_{N_{3}}\,-\,\sum_{j=1}^{k}\,p_{n_{j}}\right)\,,\;\;\;\;\;\;\;k\geq2\,,
\end{equation}
$p_{N_{3}}$ and $p_{n_{j}}$ ($j=1,\ldots,k$) being the 4-momenta of the decaying Majorana neutrino $N_{3}$ and the final state $n_{j}$, respectively. We consider the physical intermediate processes:~\footnote{Notice that the other possible cuts in Fig.~\ref{CPAsymmGraph} do not contribute to the $CP$ asymmetry as they do not correspond to physical processes.}~$N_{3}\to N_{D}+\overline{S}$ and $N_{D}+\overline{S}\to\overline{N}_{D}+S$. The corresponding tree-level amplitudes read:~\footnote{In the following
we indicate with $m_{3}$ the thermal mass of the scalar singlet $H_{3}$, which provides an infrared regulator of the 
the $N_{3}$ decay one-loop diagram.}
\begin{eqnarray}
	i\mathcal{M}^{(0)}(N_{3}\to \overline{N}_{D}+S) &=& ig\,v_{N}^{T}({\bf p}_{N})\,C^{-1}\,u_{N_{3}}({\bf p}_{N_{3}})\,,\nonumber\\
	i\mathcal{M}^{(0)}(N_{3}\to N_{D}+\overline{S}) &=& -ig\,\overline{u}_{N}({\bf p}^{\prime}_{N})\,u_{N_{3}}({\bf p}_{N_{3}})\label{ampl}\,, \\
	i\mathcal{M}^{(0)}(N_{D}+\overline{S}\to \overline{N}_{D}+S) &=&-\frac{i}{p_{H_{3}}^{2}-m_{3}^{2}}\frac{\mu^{\prime\prime}}{2}\left(\alpha^{*}\,
	v_{N}^{T}({\bf p}_{N})\,C^{-1}\, u_{N}({\bf p}^{\prime}_{N})\right)\,.\nonumber		
\end{eqnarray}
We perform the product of the three amplitudes in (\ref{ampl}) according to eq.~(\ref{CPAsymm}) and sum over the polarizations of the outgoing
fermions. After some algebra, we get
\bea
	\mathcal{M}^{(0)}(N_{3}\to \overline{N}_{D}+S)^{*}\,\mathcal{M}^{(0)}(N_{3}\to N_{D}+\overline{S})\,\mathcal{M}^{(0)}(N_{D}+\overline{S}\to \overline{N}_{D}+S)= \nonumber\\ \nonumber\\
	2\,g^{2}\,\alpha^{*}\,\frac{\mu^{\prime\prime}M^{2} M_{3}}{p_{H_{3}}^{2}-m_{3}^{2}}
	\left[\,1\,+\,\frac{\left(p_{N}\cdot p_{N}^{\prime}\right)}{M^{2}}\,+\,
	\frac{\left((p_{N}+p_{N}^{\prime})\cdot p_{N_{3}}\right)}{MM_{3}}\,\right]\,.\nonumber
\eea

\subsection*{Integration over the phase space}

The relevant  integrals in the numerator of (\ref{CPAsymm}) are
\begin{eqnarray}
	I_{n} &=& \int \frac{d^{3}p_{N}^{\prime}}{(2\pi)^{3}2E_{N}^{\prime}} \frac{d^{3}p_{S}^{\prime}}{(2\pi)^{3}2E_{S}^{\prime}}\,
			\frac{S_{n}}{p_{H_{3}}^{2}-m_{3}^{2}}\,(2\pi)^{4}\delta^{(4)}\left(p_{N_{3}}-p_{N}^{\prime}-E_{S}^{\prime}\right)\,, 		
\end{eqnarray}
where $S_{n}\in\{(p_{N}\cdot p_{N}^{\prime}),\,(p_{N}^{\prime}\cdot p_{N_{3}}),\,(p_{N}\cdot p_{N_{3}}),\,M^{2}\}$.

It is convenient to express $I_{n}$ in terms of adimensional quantities, mainly: $a\equiv E_{N}/M_{3}$, $b\equiv\left|{\bf p}_{N}\right|/M_{3}$, 
$x\equiv M/M_{3}$, $x_{S}\equiv m_{S}/M_{3}$ and $x_{3}\equiv m_{3}/M_{3}$:
\begin{eqnarray}
	I_{1} &= & \frac{1}{32\pi}\left[-2\,\kappa(1,x,x_{S})+\frac{2x^{2}-x_{3}^{2}}{b}\,C\left(x,x_{S},x_{3},a,b\right)\right]\,,\\
	I_{2,3,4} &= & \frac{1}{32\pi}\frac{1}{b}\left\{B(x,x_{S}),2a,2x^{2}\right\}\,C\left(x,x_{S},x_{3},a,b\right)\,,
\end{eqnarray}
where $\kappa$ is a kinematic factor introduced below eq.~(\ref{DecHiggs}), $B(s,t)=\sqrt{\kappa\left(1,s,t\right)^{2}+4s^{2}}$  and 
\begin{eqnarray}
	C\left(s,t,u,a,b\right) &=& \log\left(\frac{2s^{2}-u^{2}-a\,B\left(s,t\right)+b\,\kappa\left(1,t,s\right)}{2s^{2}-u^{2}-a\,B\left(s,t\right)-b\,\kappa\left(1,t,s\right)}\right)\,.
\end{eqnarray}
Now we complete the integration over the phase space in the numerator of eq.~(\ref{CPAsymm}). The relevant
integrals can be arranged in the form:
\begin{eqnarray}
	J_{n} &=& \frac{1}{4\pi}\int_{0}^{\infty}da\frac{\sqrt{a^{2}-x^{2}}}{1-a}\,I_{n}\,\delta\left(1-a-\sqrt{a^{2}-x^{2}+x_{S}^{2}} \right)\,.
\end{eqnarray}
The  full computation results in:

\begin{eqnarray}
	J_{1} &=& \frac{1}{128\pi^{2}}\left[-\kappa\left(1,x,x_{S}\right)^{2}+(2x^{2}-x_{3}^{2})\log\left(\frac{x_{3}^{2}}{x_{3}^{2}+
	\kappa\left(1,x,x_{S}\right)^{2}}\right) \right]\,,\label{J1}\\
	J_{2} \;=\; J_{3} &=&\frac{1}{128\pi^{2}}\,B\left(x,x_{S}\right)\log\left(\frac{x_{3}^{2}}{x_{3}^{2}+\kappa\left(1,x,x_{S}\right)^{2}}\right)\,,\\
	J_{4} &=& \frac{1}{64\pi^{2}}\,x^{2}\log\left(\frac{x_{3}^{2}}{x_{3}^{2}+\kappa\left(1,x,x_{S}\right)^{2}}\right)\,.
\end{eqnarray}

A similar computation applies for the denominator of eq.~(\ref{CPAsymm}). We have in this case:
\begin{equation}
	\int\,d\tilde{\Pi}_{N,S}\,\left|\mathcal{M}^{(0)}(N_{3}\to\overline{N}\,S)\right|^{2}\;=\;
	g^{2}\,\frac{M_{3}^{2}}{4\pi}\kappa\left(1,x,x_{S}\right)\left[2x+B\left(x,x_{S}\right)\right]\,.\label{denom}
\end{equation}

\mathversion{bold}
\subsection*{The $CP$ asymmetry in the decays}
\mathversion{normal}

Taking into account the results obtained in eqs.~(\ref{J1})-(\ref{denom}) and the general expression (\ref{CPAsymm}), 
we get the final expression of the $CP$ asymmetry:
\begin{equation}\label{ECPexact}
\epsilon_{CP}\;=\;
-\frac{1}{16\pi}\,\frac{\mu^{\prime\prime}}{M_{3}}\,{\rm Im}(\alpha)\,
	\frac{F_{2}(x,x_{S},x_{3})\,
	+\,2x\,\left[x+B\left(x,x_{S}\right)\right]F_{1}(x,x_{S},x_{3})}
	{2x+B\left(x,x_{S}\right)}\,,
\end{equation}
where 
\begin{eqnarray}
	F_{1}(x,x_{S},x_{3}) &=& \frac{1}{\kappa\left(1,x,x_{S}\right)}\log\left(\frac{x_{3}^{2}}{x_{3}^{2}+\kappa\left(1,x,x_{S}\right)^{2}}\right)\,,\\
	F_{2}(x,x_{S},x_{3}) &=& -\kappa\left(1,x,x_{S}\right)+(2x^{2}-x_{3}^{2})F_{1}(x,x_{S},x_{3})\,.
\end{eqnarray}
Therefore in the limit $m_{S}$, $M\ll M_{3}$, which we are interested in, we get 
the approximation reported in eq.~(\ref{ApprECP}):
\begin{equation}
	\epsilon_{CP}\;\simeq\;-\frac{1}{16\pi}\,\frac{\mu^{\prime\prime}}{M_{3}}\,{\rm Im}(\alpha)\,.Ê\nonumber 
\end{equation}

\section{Boltzmann Equations}\label{BESec}

In this appendix, we introduce the set of Boltzmann Equations (BE) that are used for the numerical evaluation of the baryon asymmetry. More details on the network of BE can be found in the appendices of~\cite{Davidson:2008bu} and \cite{GiudiceLepto}. For a given particle (asymmetry) $X$, we denote as usual by $Y_{X}$ its comoving number density, \ie \, the number density normalized to the entropy density. We assume Maxwell-Boltzmann statistics for both fermions and scalars.
In an expanding Universe, the evolution of $Y_{X}$ is governed by the Boltzmann equation:
\bea
s\,H(z)\frac{ d Y_X}{dz}\;= -\sum_{a,i,j, \etc} \left \lbrack X\,a\, \rightleftarrows i\,j\, \right\rbrack \,,\nonumber
\eea
where 
\bea
\left\lbrack X\,a\, \rightleftarrows i\,j\,\right\rbrack \equiv \frac{Y_{X}}{Y_{X}^{eq}}\frac{Y_{a}}{Y_{a}^{eq}}\, \gamma^{eq}(X\,a\to i\,j)-\frac{Y_{i}}{Y_{i}^{eq}}\frac{Y_{j}}{Y_{j}^{eq}}\, \gamma^{eq}(i\,j\to X\,a) \label{eveq} \,, \nonumber
\eea
and $z=M_{3}/T$ is the evolution parameter, while $\gamma^{eq}$ are the equilibrium reaction densities of the different processes. We will limit our analysis to $1\leftrightarrow 2$ and $2\leftrightarrow 2$ processes, but will include the on-shell part of some $2\leftrightarrow 3$ scatterings for consistency.
If these processes conserve $CP$, then we use the notation $\left[X\,a \leftrightarrow i\,j\right]$, as $\gamma^{eq}(X\,a\to i\,j)=\gamma^{eq}(i\,j\to X\,a)$.

In a radiation dominated Universe, the Hubble constant $H(T)$ and the entropy density $s$ are given by
\bea
	H(T)&=& \sqrt{\frac{4\,\pi^{3}\,g_{*}}{45}}\frac{T^{2}}{M_{\rm pl}}\,, \quad s=g_{*}\,\frac{2\,\pi^2}{45}\,T^{3}\,. \nonumber
\eea
In these equations, $g_{*}$ is the number of relativistic degrees of freedom present in the thermal bath at the leptogenesis time scale.
In the case of the SM, at temperatures above the EWSB, one has $g_{*}^{\rm SM}=106.75$. 
Assuming that the non-SM scalars $S$, $H_{2}$ and $H_{3}$ and the Dirac fermion $N_{D}$ are relativistic particles
at $T\approx M_{3}$, we obtain: $g_{*}=g_{*}^{\rm SM}+46/4=118.25$. 

As already explained in Section~\ref{2stepslep}, the main source of $N_{D}$ and $S$ asymmetry production during the first stage of leptogenesis are the $CP$-violating decays and inverse decays of $N_{3}$, 
\bea
\gamma^{eq}\left(N_3\rightleftarrows N_D\,\ovl{S} \right)\equiv\gamma_{D}\left(\frac{1\pm\eps_{CP}}{2} \right)= \gamma^{eq}\left(N_3\leftrightarrows \ovl{N}_D\, S \right)\,,\nonumber
\eea
where $\epsilon_{CP}$ is the $CP$ asymmetry in the decays, defined in eq.~(\ref{CPAsymm}), 
and $\gamma_{D}$ is the $CP$ conserving total decay width of $N_{3}$. The last equality results from $CPT$ invariance.\\
We further include in the BE $\D N_{D}=\D S=2$ scatterings shown in Fig.~\ref{DN2scatt}, whose corresponding collision rates are denoted as:
\bea
	\gamma^{eq}(N_D\,\ovl{S}\rightleftarrows \ovl{N}_D\,S)\equiv\gamma_{\Delta 2}^{a}\quad {\rm and}\quad	\gamma^{eq}(N_{D}\,N_{D}\leftrightarrow S\,S)\equiv\gamma_{\D 2}^{b}\,.\nonumber
\eea
Note that, as in standard leptogenesis, $N_D\,\ovl{S}\rightleftarrows \ovl{N}_D\,S$ processes mediated by $N_{3}$ in a $s$-channel develop an on-shell part, which is $CP$-violating. To avoid double-counting of this resonant part, already accounted for by the inverse decays, the on-shell contribution should be subtracted from the full $N_D\,\ovl{S}\leftrightarrow \ovl{N}_D\,S$ scattering rate.

In addition to the standard source term given by the decays of $N_{3}$, we include the $CP$ violation arising from the $2\leftrightarrow2$ scatterings involving an external $N_{3}$, which also  depends on the $CP$-violating phase $\al$ entering in $\eps_{CP}$, eq.~(\ref{ApprECP}). The $CP$ asymmetry for each diagram is computed as
in the standard leptogenesis scenario, $e.g.$  \cite{Abada:2006ea} and \cite{NardiCP}. 
However, in our model a contribution to $CP$ asymmetry in the $N_{3}$-scatterings arises  from both $s$-, $t$- and $u$-channels, as depicted in Fig.~\ref{DN1scatt}.
The corresponding thermal rates, in this case, are: 

\bea
\rm{a)} && \gamma^{eq}(N_D\,\,N_3\rightleftarrows H_3\,\,\ovl{S})\;\equiv\; \gamma_{N_3}^{a} \left(1\pm\eps_{CP}^{a}\right)\,=\,\gamma^{eq}(\ovl{N}_D\,\,N_3\leftrightarrows \ovl{H}_3\,\,S)\,.\nonumber\\
\rm{b)} &&\gamma^{eq}(N_3\,\,S\rightleftarrows \ovl{N}_D\,\,H_3)\equiv\gamma_{N_3}^{b} \left(1\mp\eps_{CP}^{b}\right)\,=\,\gamma^{eq}(N_3\,\,\ovl{S}\leftrightarrows N_D\,\,\ovl{H}_3)\,.\nonumber\\
\rm{c)} &&\gamma^{eq}(N_D\,\,S\rightleftarrows N_3\,\,H_3)\;\equiv\;\gamma_{N_3}^{c} \left(1\mp\eps_{CP}^{c}\right)\,=\gamma^{eq}(\ovl{N}_D\,\,\ovl{S}\leftrightarrows  N_3\,\,\ovl{H}_3)\,.\nonumber\\
\rm{d)} &&\gamma(N_3\,\,S\rightleftarrows H_1\,\,\ell)\;\equiv\;\gamma_{N_3}^{d} \left(1\pm\eps_{CP}\right)\,=\gamma(N_3\,\,\ovl{S}\leftrightarrows \ovl{H}_1\,\,\ovl{\ell})\,.\nonumber\\
\rm{e)} &&\gamma(N_3\,\,S\rightleftarrows \ovl{H}_2\,\,\ovl{\ell})\;\equiv\; \gamma_{N_3}^{e} \left(1\pm \eps_{CP}\right)\,=\gamma(N_3\,\,S\leftrightarrows H_2\,\,\ell)\,.\nonumber
\eea
The $CP$ asymmetries in the scattering, $\eps_{CP}^{k}$ ($k=a,\etc,e$), are defined by $\eps_{CP}^{k} \equiv  \eps_{CP}\,\D K^{k}$, with
\bea
		\D K^{a,b} \equiv\frac{\left(\gamma_{N_{3}}^{a,b}\right)_{t}-\left(\gamma_{N_{3}}^{a,b}\right)_{s}}{\gamma_{N_{3}}^{a,b}}\,,\quad
		 \D K^{c} \equiv\frac{\left(\gamma_{N_{3}}^{c}\right)_{t}-\left(\gamma_{N_{3}}^{c}\right)_{u}}{\gamma_{N_{3}}^{c}}\, , \quad 
		 \D K^{d}=\D K^{e}=1\,. \nonumber
\eea
Here $\left(\gamma_{N_{3}}^{k}\right)_{c}$, $c=(s,t,u)$, corresponds to the $s$-, $t$- and $u$-channels of the different processes shown in Fig.~\ref{DN1scatt}.
Notice that, similarly to the $\D N_{D}=\D S=2$ scatterings considered before, as explained in \cite{NardiCP},
we have to subtract the resonant $CP$-violating contribution of the $2\leftrightarrow3$  processes in which $N_{3}$ is 
exchanged in s-channel. The non-resonant parts of such processes
are not taken into account in our computation, since they are at higher order in the couplings.

As regards $N_{3}$ three-body decays, which are at the same order in the couplings as $2\leftrightarrow2$ scatterings on $N_{3}$, they are phase-space suppressed and so give a sub-leading contribution with respect to the two-body decays \cite{NardiCP}, and we consequently do not include them.

We further consider the effect of $S$ self-annihilations (see Fig.~\ref{SSelfAnn}), which
could washout the asymmetry $Y_{\D S}$ for large values of 
the coupling $h$ (see eq.~(\ref{VDM})). The related interaction density rate is noted
\bea 
\gamma^{eq}(S\,S\,\leftrightarrow\,H_{1}\,\ovl{H}_{2})\equiv\gamma_{SS}\,. \nonumber
\eea

Several processes participate in the second phase of leptogenesis. Besides the scatterings on $N_{3}$, the $\gamma_{N_{3}}^{d,e}$ discussed above, we 
include the following interactions, at the lowest order in the neutrino Yukawa couplings: 
\begin{itemize}
\item[a)] $N_{D}$ decays and inverse decays: $\gamma^{eq}(N_{D}\,\leftrightarrow\,\ell_{\alpha}\,H_{1})\equiv\gamma_{D\ell}$ and
		$\gamma^{eq}(N_{D}\,\leftrightarrow\,\ovl{\ell_{\beta}}\,\ovl{H}_{2})\equiv\gamma_{D\ovl{\ell}}$.
\item[b)] $\Delta L=1$, $H_1$-mediated scatterings with top-quark:~$\gamma^{eq}(N_D \ovl{\ell} \leftrightarrow Q_{3} \, \ovl{t})\equiv\gamma_{N_D}^{s}$ for the s-channel  contribution and $\gamma^{eq}(N_D q_{3} \leftrightarrow \ell \, t)+\gamma^{eq}(N_D \ovl{t} \leftrightarrow \ell \,\ovl{q_{3}})=2\,\gamma_{N_D}^{t}$ for the t-channel contributions.	 	 
\end{itemize}
As already stated in Section~\ref{2stepslep}, the leptons participate in $N_{D}$ mediated $\D L =2$ scatterings: 
$\gamma_{\ell\ell}^{a}\equiv\gamma^{eq}(\ell\, H_1 \leftrightarrow \ovl{\ell}\,\ovl{H}_{2})$ and $\gamma_{\ell\ell}^{b}\equiv\gamma^{eq}(\ell\,\ell \leftrightarrow \ovl{H}_{1}\, \ovl{H}_{2})$.

We are ready now to report the complete set of Boltzmann equations relevant for the computation of the baryon asymmetry of the Universe
in the two-step leptogenesis scenario described in the text.
We include all the interaction terms introduced above and we use the simplified notation 
$y_{N_{3}}\equiv Y_{N_{3}}/Y_{N_{3}}^{eq}$, $y_{\Delta X}\equiv Y_{\Delta X}/Y_{X}^{eq}$ and 
$Y^{\prime}_{X}\equiv(s\,H(z)) \,d Y_{X}/dz$. At first order in the asymmetry (zeroth order for $N_{3}$), the full system of Boltzmann equations
is the following:
\bea
Y_{N_{3}}^{\prime}&=&-\left[N_{3} \rightleftarrows N_{D}\,\ovl{S}\right]-\left[N_{3} \rightleftarrows \ovl{N}_{D}\,S\right]-\left[N_{D}\,N_{3} \rightleftarrows H_{3}\,\ovl{S}\right]-\left[\ovl{N}_{D}\,N_{3} \rightleftarrows \ovl{H}_{3}\,S\right] \nonumber \\
&-&\left[S\,N_{3} \rightleftarrows \ovl{N}_{D}\,H_{3}\right]-\left[\ovl{S}\,N_{3} \rightleftarrows N_{D}\,\ovl{H}_{3}\right]+\left[S\, N_{D}\rightleftarrows N_{3}\,H_{3}\right]+\left[\ovl{S}\, \ovl{N}_{D}\rightleftarrows N_{3}\,\ovl{H}_{3}\right] \nonumber \\
&+&\left[\ell \,H_{1} \rightleftarrows N_{3}\,S\right]+\left[\ovl{\ell} \,\ovl{H}_{1} \rightleftarrows N_{3}\,\ovl{S}\right]+\left[\,H_{2}\,\ell \rightleftarrows N_{3}\,\ovl{S}\right]+\left[\,\ovl{H}_{2}\,\ovl{\ell} \rightleftarrows N_{3}\,S\right] \,,\nonumber
\eea
\bea
Y_{N_D}^{\prime}&=& \left[N_{3} \rightleftarrows N_{D}\,\ovl{S}\right]-\left[ N_{D}\,\ovl{S}\rightleftarrows \ovl{N}_{D}\,S\right]-\left[N_{D}\,N_{D}\leftrightarrow S\,S\right]\nonumber \\
&-&\left[N_{D}\,N_{3} \rightleftarrows H_{3}\,\ovl{S}\right]-\left[N_{D}\,\ovl{H}_{3} \rightleftarrows N_{3}\,\ovl{S}\right]-\left[N_{D}\,\ovl{S}\rightleftarrows N_{3}\,H_{3}\right] \nonumber \\
&-&\left[N_{D} \leftrightarrow \ell H_{1}\right]-\left[N_{D} \leftrightarrow \ovl{\ell}\,\ovl{H}_{2}\right]-\left[N_{D}\,\ovl{\ell} \leftrightarrow q_{3}\,\ovl{t}\right]-\left[N_{D}\,q_{3}\leftrightarrow \ell \,t\right]-\left[N_{D}\,\ovl{t} \leftrightarrow \ell\,\ovl{q_{3}}\right]\,,\nonumber
 \eea
\bea
Y_{S}^{\prime}&=&\left[N_{3} \leftrightarrows S\,\ovl{N}_{D}\right]-\left[S\,\ovl{N}_{D}\rightleftarrows \ovl{S}\,N_{D} \right]-\left[S\,S\,\leftrightarrow N_{D}\,N_{D}\right]\nonumber \\
&-&\left[S\,\ovl{H}_{3} \rightleftarrows N_{3}\,N_{D}\right]-\left[S\,N_{3} \rightleftarrows \ovl{N}_{D}\,H_{3}\right]-\left[S\, N_{D}\rightleftarrows N_{3}\,H_{3}\right]\nonumber \\
&-&\left[S\,N_{3}\rightleftarrows \ell H_{1}\right]-\left[S\,N_{3}\rightleftarrows \ovl{\ell}\,\ovl{H}_{2}\right]-\left[S\,S\leftrightarrow H_{1}\,\ovl{H}_{2}\right]\,, \nonumber 
\eea
\bea
Y_{\ell}^{\prime}&=& \left[N_{D}\leftrightarrow \ell H_{1}\right]+\left[\ovl{N}_{D}\leftrightarrow \ell H_{2}\right]-\left[\ell\,\ovl{N}_{D}\leftrightarrow \ovl{q_{3}}\,t\right]-\left[\ell\,t \leftrightarrow N_{D}\,q_{3}\right]-\left[\ell \, \ovl{q_{3}}\leftrightarrow N_{D}\,\ovl{t}\right]\nonumber \\
&-&\left[\ell \,H_{1} \rightleftarrows N_{3}\,S\right]-\left[\ell \,H_{2}\rightleftarrows N_{3}\,\ovl{S}\right]-\left[\ell\,H_1 \leftrightarrow \ovl{\ell}\,\ovl{H}_{2}\right]
-\left[\ell\,\ell \leftrightarrow \ovl{H}_{1}\, \ovl{H}_{2}\right]\,, \nonumber
\eea
\bea
Y_{H_{1}}^{\prime} &=& \left[N_{D} \leftrightarrow H_{1}\,\ell\right]-\left[H_{1}\,\ell \rightleftarrows N_{3}\,S\right]-\left[H_{1}\,\ovl{H}_{2} \leftrightarrow \,S\,S\right]
-\left[\ell\,H_1 \leftrightarrow \ovl{\ell}\,\ovl{H}_{2}\right]-\left[H_{1}\, H_{2} \leftrightarrow  \ovl{\ell}\,\ovl{\ell}\right]\,,
\nonumber 
\eea
\bea
Y_{H_{2}}^{\prime} &=& \left[N_{D} \leftrightarrow \ell\,H_{2}\right]-\left[\,H_{2}\,\ell \rightleftarrows N_{3}\,\ovl{S}\right]-\left[H_{2}\,\ovl{H}_{1} \leftrightarrow \,\ovl{S}\,\ovl{S}\right]
-\left[\ell\,H_2 \leftrightarrow \ovl{\ell}\,\ovl{H}_{1}\right]-\left[H_{1}\, H_{2} \leftrightarrow  \ovl{\ell}\,\ovl{\ell}\right]\,, \nonumber  
\eea
\bea
Y_{H_{3}}^{\prime} &=& -\left[H_{3}\,\ovl{S}Ê\rightleftarrows N_{3}\,N_{D}\right]-\left[H_{3}\,\ovl{N}_{D} \rightleftarrows N_{3}\,S\right]-\left[H_{3}\,N_{3} \rightleftarrows N_{D}\,S\right]\,. \nonumber
\eea
The evolution equations of the antiparticles are obtained by taking the $CP$ conjugates of the different rates. The Boltzmann equations of the number density (asymmetry) finally read:
 \bea
Y_{N_{3}}^{\prime}& =& \left(1-y_{N_{3}}\right) \left( \gamma_D + 2 \sum_{k=a,\ldots,e} \gamma_{N_3}^{k} \right) \,,\\
Y_{\Delta N_D}^{\prime}&=&
\left(y_{N_{3}}-1\right)\left(\eps_{CP}\,\gamma_{D}+2\,\eps_{CP}^{a}\,\gamma_{N_{3}}^{a}-2\,\eps_{CP}^{b}\,\gamma_{N_{3}}^{b}-2\,\eps_{CP}^{c}\,\gamma_{N_{3}}^{c}\right) \nonumber \\
&-& 2 \left( \gamma_{\D\,2}^{a}+2\gamma_{\D\,2}^{b} \right)\left( \YDn - \YDs \right)
-\gamma_{N_{3}}^{a}\left(y_{N_{3}}\YDn-\YDhc+\YDs\right)\nonumber\\
&-&\gamma_{N_{3}}^{b}\left(\YDn-\YDhc+y_{N_{3}}\YDs\right)-\gamma_{N_{3}}^{c}\left(\YDn-y_{N_{3}}\YDhc+\YDs\right)\,\nonumber \\
&-&\gamma_{D\ell}\left(\YDn-\YDl-\YDha\right)-\gamma_{D\ovl{\ell}}\left(\YDn+\YDl+\YDhb \right)\nonumber\\
&-&\left( \gamma_{N_D}^{s} + 2\gamma_{N_3}^{t} \right)\left(\YDn-\YDl \right)\,,\\
\label{BE:YDN}
Y_{\Delta S}^{\prime}&=& -\left(y_{N_{3}}-1\right)\left(\eps_{CP}\,\gamma_{D}+2\,\eps_{CP}^{a}\,\gamma_{N_{3}}^{a}-2\,\eps_{CP}^{b}\,\gamma_{N_{3}}^{b}-2\,\eps_{CP}^{c}\,\gamma_{N_{3}}^{c}+2\,\eps_{CP}\,\gamma_{N_{3}}^{d}+2\,\eps_{CP}\,\gamma_{N_{3}}^{e}\right) \nonumber \\
&-&2 \left( \gamma_{\D\,2}^{a}+2\gamma_{\D\,2}^{b} \right)\left(\YDs-\YDn \right) \,-\,2\gamma_{SS}\left(2\YDs-\YDha+\YDhb\right)
 \nonumber\\
&-&\gamma_{N_{3}}^{a}\left(\YDs-\YDhc+y_{N_{3}}\YDn\right)-\gamma_{N_{3}}^{b}\left(y_{N_{3}}\YDs-\YDhc+\YDn\right)\nonumber \\
&-&\gamma_{N_{3}}^{c}\left(\YDs-y_{N_{3}}\YDhc+\YDn\right)-\gamma_{N_{3}}^{d}\left(y_{N_{3}}\YDs-\YDha-\YDl\right)\nonumber \\
&-&\gamma_{N_{3}}^{e}\left(y_{N_{3}}\YDs+\YDhb+\YDl\right)\,, \label{BE:YDS}
\eea

\bea
Y_{\Delta \ell }^{\prime}&=& -\left(y_{N_{3}}-1\right)\left(2\,\eps_{CP}\,\gamma_{N_{3}}^{d}-2\,\eps_{CP}\,\gamma_{N_{3}}^{e}\right)- \gamma_{D\,\ell} \left(\YDl +\YDha - \YDn\right)\nonumber\\ 
&-& \gamma_{D\,\ovl{\ell}} \left( \YDl +\YDhb + \YDn \right) +\left( \gamma_{N_D}^{s} + 2\gamma_{N_3}^{t} \right)\left(\YDn-\YDl \right)
-\,\left(\gamma_{\ell\ell}^{a}+2\,\gamma_{\ell\ell}^{b}\right)\left(2\YDl+ \YDha+\YDhb\right)\nonumber\\
&-&\gamma_{N_{3}}^{d}\left(\YDha+\YDl-y_{N_{3}}\YDs\right) -\gamma_{N_{3}}^{e}\left(\YDhb+\YDl+y_{N_{3}}\YDs\right)\,,\\
Y_{\Delta H_1 }^{\prime}&=&-2\,\left(y_{N_{3}}-1 \right)\,\eps_{CP}\, \gamma_{N_3}^{d} -\gamma_{D\,\ell} \left(\YDha + \YDl - \YDn\right)
-\left(\gamma_{\ell\ell}^{a}+\gamma_{\ell\ell}^{b}\right)\left(2\YDl+ \YDha+\YDhb\right) \nonumber \\&-&\gamma_{N_{3}}^{d}\left(\YDha+\YDl-y_{N_{3}}\YDs\right) -\gamma_{SS}\left(\YDha - \YDhb-2\,\YDs\right)\,, \\
Y_{\Delta H_2 }^{\prime}&=&2\,\left(y_{N_{3}}-1 \right)\,\eps_{CP}\, \gamma_{N_3}^{e} - \gamma_{D\,\ovl{\ell}} \left(\YDhb + \YDl + \YDn \right)
-\left(\gamma_{\ell\ell}^{a}+\gamma_{\ell\ell}^{b}\right)\left(2\YDl+ \YDha+\YDhb\right) \nonumber \\ 
&-& \gamma_{N_3}^{e} \left(\YDhb + \YDl + \YDs\, \Yn \right)\-\gamma_{SS}\left(\YDhb-\YDha+2\,\YDs\right)\,, \\
Y_{\Delta H_3 }^{\prime}&=&\left(y_{N_{3}}-1 \right) \left( 2\,\eps_{CP}^{a}\,\gamma_{N_{3}}^{a}-2\,\eps_{CP}^{b}\,\gamma_{N_{3}}^{b} -2\,\eps_{CP}^{c}\,\gamma_{N_{3}}^{c}\right) 
 -\gamma_{N_3}^{a}\left(\YDhc - \YDs - \YDn \Yn\right)\nonumber\\&-&\gamma_{N_3}^{b} \left( \YDhc - \YDn - \YDs \Yn\right)- \gamma_{N_3}^{c} \left(\YDhc \Yn -\YDn - \YDs  \right)\,. 
 \eea

\section{Chemical Equilibrium Conditions}\label{ChemEqCon}

\begin{table}[ht]
\begin{center}
\begin{tabular}{|c|c|c|c||c|c||c|}
\hline \hline
    & $c_{B-L}$ & $c_{N_{D}}$ & $c_{S}$ & $c_{B-L}$ & $c_{N_{D}}$ & $c_{B-L}$\\[3mm]
  $\mu_{H_{1}}$ & $\frac{1}{16}$& $\frac{15}{16}$ & $0$  & $\frac{1}{16}$ & $\frac{15}{16}$ & $-\frac{1}{14}$ \\[3mm]
  $\mu_{H_{2}}$ & $-\frac{11}{16}$ & $-\frac{101}{16}$ & $0$ & $-\frac{11}{16}$ & $-\frac{101}{16}$ & $\frac{3}{14}$\\[3mm]
  $\mu_{H_{3}}$ & $-\frac{9}{8}$ & $-\frac{101}{8}$ & $-\frac{1}{2}$ & $-\frac{21}{16}$ & $-\frac{231}{16}$ & $\frac{3}{4}$ \\[3mm]
  $\mu_{S}$ & $0$ & $0$ & $1$ & $\frac{3}{8}$ & $\frac{29}{8}$ & $-\frac{1}{7}$ \\[3mm]
  $\mu_{N_{D}}$ & $0$ & $1$ & $0$ & $0$ & $1$ & $-\frac{1}{7}$ \\[3mm]
  $\mu_{\ell}$ & $-\frac{1}{16}$ & $\frac{1}{16}$ & $0$ & $-\frac{1}{16}$ & $\frac{1}{16}$ & $-\frac{1}{14}$ \\[3mm]
\hline\hline
\end{tabular}
\caption{Chemical equilibrium coefficients in eq.~(\ref{chcoeff}) for the three cases discussed in the text.}
\end{center}
\label{ChemRel}
\end{table}%

We derive in this section the chemical equilibrium conditions 
provided by all the interactions which are in equilibrium
at the leptogenesis epoch, $T\sim M_{3}\lesssim 10^{5\div6} \GeV$.

The chemical potentials of each generation of $SU(2)_{W}$ quark doublets, $Q_{i}$, and singlets, $u_{Ri}$ and $d_{Ri}$, 
are denoted by $\mu_{Q_{i}}\equiv\mu_{Q}$, $\mu_{u_{Ri}}\equiv \mu_{u}$ and  $\mu_{d_{Ri}}\equiv\mu_{d}$, respectively. 
Concerning the lepton fields, we define for each flavor $\alpha$ the corresponding chemical potentials as:
$\mu_{\ell_{\alpha}}\equiv \mu_{\ell}$, $\mu_{e_{R\alpha}}\equiv\mu_{e}$. We denote with $\mu_{N}$ the
chemical potential of $N_{D}$. Analogously,   
for each scalar field in the model we define, in a consistent notation: $\mu_{H_{1,2,3}}$ and $\mu_{S}$. We remark that
the chemical potentials of the SM fermions are assumed to be independent of the generation index, because of the
the rapid flavor mixing interactions which occur at the leptogenesis time \cite{Harvey}. 

The number density asymmetries are related to the particle chemical potentials
through the relations:
\bea
	Y_{\D X} & \simeq & \frac{g_{X}\, T^{3}}{3s}\,\mu_{X}\quad\,{\rm for\;\;bosons}\,,\\
	Y_{\D X} & \simeq & \frac{g_{X}\, T^{3}}{6s}\,\mu_{X}\quad\,{\rm for\;\;fermions}\,,
\eea
where $g_{X}$ is the number of internal degrees of freedom of the particle $X$.
The total baryon and lepton number asymmetries can be expressed in terms of the fermion
chemical potentials:
\begin{eqnarray}
&&	Y_{\Delta B}=\frac{T^{3}}{2s}\left(2\mu_{Q}+\mu_{u}+\mu_{d}\right)\,,\quad  Y_{\Delta L}=\frac{T^{3}}{2 s}\left(2\mu_{\ell}+\mu_{e}+\frac{2}{3}\mu_{N_{D}} \right)\,.
\end{eqnarray}
Taking into account the definitions given above, we have the following relations \cite{Harvey}:
\begin{itemize}
\item[1.] QCD and  $SU(2)_{W}$ sphaleron interactions:
\begin{eqnarray}\label{start}
	&& 2\mu_{Q}-\mu_{u}-\mu_{d}=0\,,\\
	&& 3\mu_{Q}+\mu_{\ell}=0\,.
\end{eqnarray}
\item[2.] Hypercharge neutrality:
\begin{eqnarray}
&&	3\left(\mu_{Q}+2\mu_{u}-\mu_{d}-\mu_{\ell}-\mu_{e}\right)+2\left(\mu_{H_{1}}+\mu_{H_{2}}\right)=0\,.	
\end{eqnarray}
\item[3.] Charged lepton Yukawa interactions:
\begin{eqnarray}
&&	\mu_{\ell}-\mu_{H_{1}}-\mu_{e}=0\,,\\
&&	\mu_{Q}+\mu_{H_{1}}-\mu_{u}=0\,,\\
&&	\mu_{Q}-\mu_{H_{1}}-\mu_{d}=0\,.
\end{eqnarray}
\item[4.] Lepton number conserving Dirac neutrino Yukawa interactions:
\begin{eqnarray}
&&	\mu_{N_{D}}-\mu_{H_{1}}-\mu_{\ell}=0\,.
\end{eqnarray}
\item[5.] $(B-\widetilde{L})$ conservation:
\begin{eqnarray}
&&	3\,\left(2\mu_{Q}+\mu_{u}+\mu_{d}\right)-3\left(2\mu_{\ell}+\mu_{e}\right)-2\mu_{N_{D}}-2\left(\mu_{S}+2\mu_{H_{3}}-4\mu_{H_{2}}\right) =0\,.\label{end}	
\end{eqnarray}
\end{itemize}
We notice that the QCD sphaleron condition is redundant in this case, as all quark Yukawa
interactions are in equilibrium. 

The different chemical equilibrium conditions enforce relations among the chemical potentials, which then can be expressed in terms of a subset of them. We set
\bea
	\mu_{X}&=& c_{B-L}\,\mu_{B-L}\,+\,c_{N_{D}}\,\mu_{N_{D}}\,+\,c_{S}\,\mu_{S}\,,\label{chcoeff}
\eea
where we define $\mu_{B-L}$ through the relation: $Y_{\D(B-L)}\equiv Y_{\D B}-Y_{\D L}\equiv \mu_{B-L} \,T^{3}/(2s)$.
We then distinguish three possible scenarios:
\begin{itemize}
	
	\item[A.] Lepton number violating neutrino Yukawa interactions and $S$ self-annihilation are decoupled,
	which corresponds to the set of equilibrium conditions $1-5$ listed above. The different chemical potentials can be expressed in terms of the set $(\mu_{B-L},\mu_{N_{D}},\mu_{S})$.
	
	\item[B.] $S$ self annihilations are always in equilibrium, but lepton number violating Yukawa
	interactions are still decoupled. An additional equilibrium condition is enforced: 
	\bea
	2\mu_{S} -\mu_{H_{1}}+\mu_{H_{2}}=0\,.\label{chS2}
	\eea
	Only two chemical potentials are independent, that we choose to be $(\mu_{B-L},\mu_{N_{D}})$.
	
	\item[C.] All interactions listed above, as well as lepton number violating Yukawa interactions, are in thermal equilibrium during the leptogenesis era:
	\bea
	\mu_{N_{D}}+\mu_{H_{2}}+\mu_{\ell}=0\,.\label{chy2}
	\eea
	In this case, all chemical potentials are proportional and can be expressed for example in terms of $\mu_{B-L}$.

\end{itemize}
The coefficients $c_{X}$ in eq.~(\ref{chcoeff}), corresponding to the three cases listed above
are reported in Tab.~\ref{ChemRel}.
In the first two cases the final baryon asymmetry is given by
\bea
	Y_{\D B} &=& \frac{1}{4}\,Y_{\D(B-L)}-\frac{1}{8}\,Y_{\D N_{D}}\,.\label{YBc12}
\eea 
In the last scenario, which corresponds to the case discussed in Section~\ref{2stepslep}, where all the interactions listed above are in thermal equilibrium
during the generation of the BAU, we have:
\bea
	Y_{\D B} &=& \frac{2}{7}\,Y_{\D(B-L)}\,.\label{YBc3}
\eea
Notice that expressions (\ref{YBc12}) and (\ref{YBc3}) should be considered valid up to the decoupling of $N_{D}$,
$i.e.$ for $\Gamma_{N_{D}}\gg H$.

\end{document}